\documentclass[epjST]{svjour}
\usepackage{graphicx,amssymb,amsmath}
\begin{document}
\title{CMS results and status}
\author{Lars Sonnenschein\thanks{\email{Lars.Sonnenschein@cern.ch}}
\\
on behalf of the CMS collaboration 
}
\institute{RWTH Aachen University, III. Phys. Inst. A}
%
\abstract{
The CMS experiment is a multi-purpose detector successfully operated at the LHC
where predominantly $pp$ collisions take place at various centre of mass energies
up to $\sqrt{s}=8$~TeV at present. Discussed are $pp$ collision results until end 
of 2011, corresponding to centre of mass energies of up to $\sqrt{s}=7$~TeV.
The excellent performance of the accelerator 
and the experiment allows for dedicated physics measurements over a wide range
of subjects, starting from particle identification, encompassing 
Standard Model measurements in multijet, boson, heavy flavour and top quark 
physics, building the basis for new physics searches interpreted within the framework
of various models and theories.
} 
\maketitle
\section{Introduction} \label{intro}
The performance of the Compact Muon Solenoid (CMS) experiment~\cite{cms} is discussed 
in detail, followed by measurements dedicated to give a more accurate account
of the Standard Model (SM) by means of
multijet and boson production, followed by a rich $B$ physics programme 
including the observation of a new resonant state $\Xi_b^{*0}$ and setting 
severe constraints on new physics by means of searches for 
$B_{(d,s)}^0\rightarrow \mu^+\mu^-$ decays. 
Furthermore the heaviest Standard Model candle - the top quark -
is produced copiously at the Large Hadron Collider (LHC), providing the opportunity 
for detailed studies
of $t\bar{t}$ and single top production cross sections as well as measurements 
of top quark properties, most prominently its mass, which allows 
together with precision measurements of the $W$ boson mass to set indirectly 
constraints on the Brout (alias Higgs) boson mass. Various search channels
are discussed and combined. The search for supersymmetry is exemplarily
elaborated for hadronic, semi-leptonic and leptonic final states, followed by 
a summary of excluded mass scales in simplified model spectra and exclusion
limits in the phase space of the constrained Minimal Supersymmetric SM (cMSSM).
Finally exotic searches are presented, encompassing searches for TeV gravity
and various other models in lepton, lepton plus jet and jet final states.

Throughout this article the convention $c \equiv 1$ is adopted for the speed of light.

\begin{figure}[b]
 \vspace*{-12ex}
 \resizebox{0.5\columnwidth}{!}{%
 \hspace*{-4ex} \includegraphics{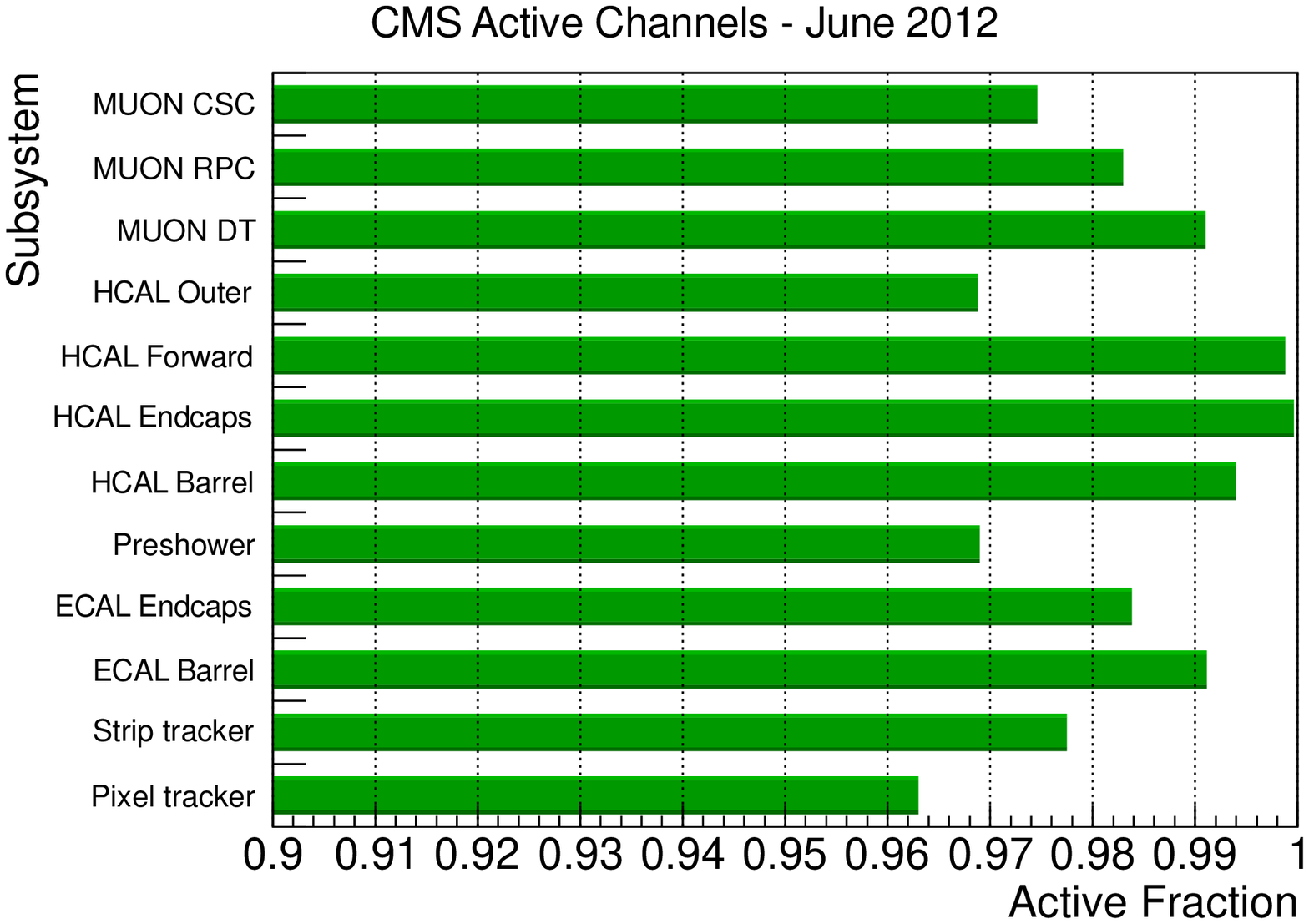} 
 \hspace*{-4ex}
}
 \unitlength 1cm
 \begin{picture}(10.,6.)
 \put(-0.5,4.9){ \includegraphics[width=5.2cm, angle=270]{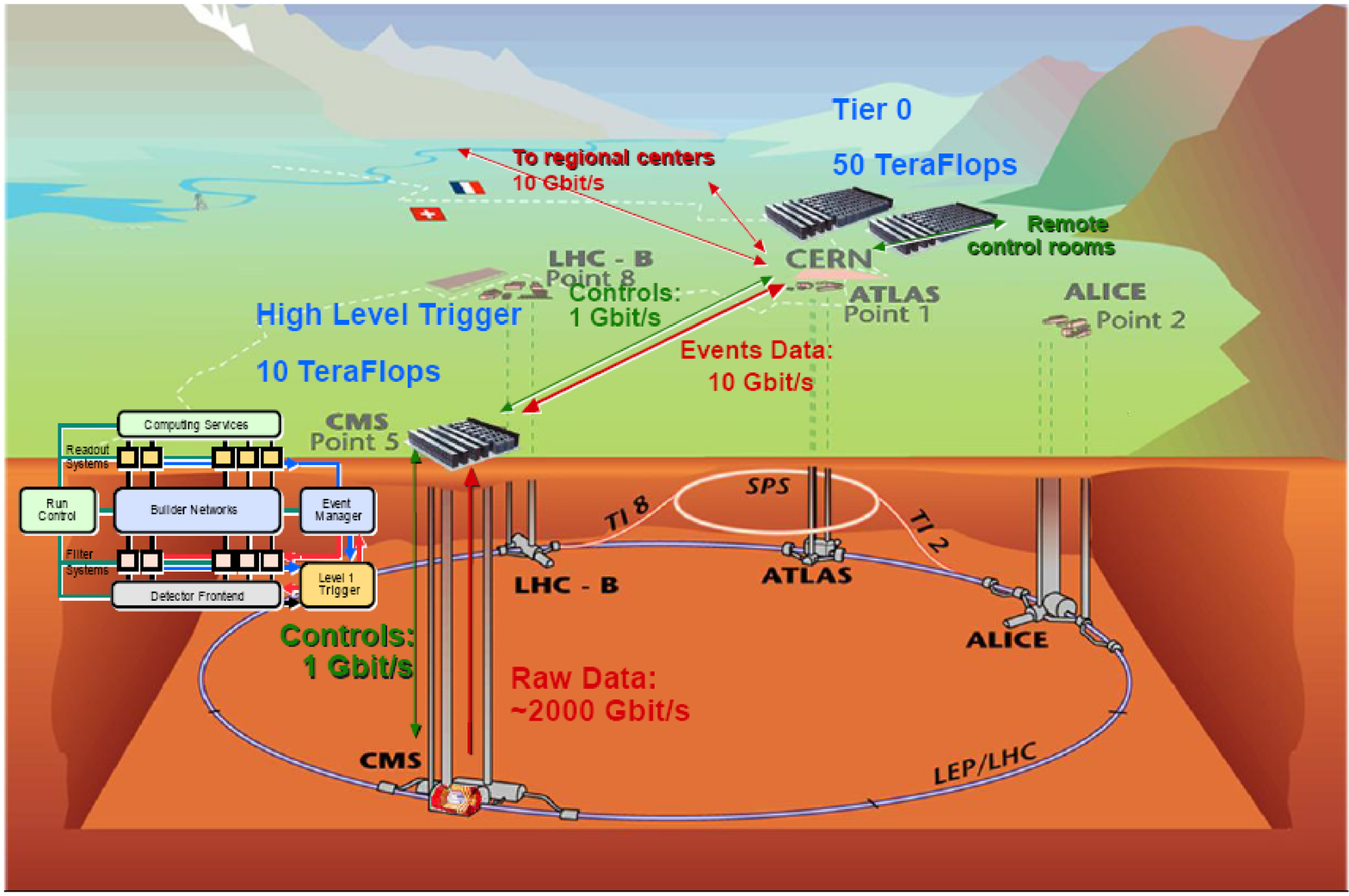} }
\end{picture}

\caption{ \label{fig0:CMSperformance}
Active fraction of the CMS subdetectors (left). In the average 98\% of the detector is
operational. On the right is shown an overview of the LHC experiments, 
in particular the CMS experiment with its infrastructure and data flow.
}
\end{figure}

\section{CMS status and operation}
The CMS~\cite{cms} experiment is a multi-purpose detector operated a the 
LHC at CERN. In the barrel part of the detector concentric around 
the beam pipe are installed from inside to outside three layers of silicon pixels,
surrounded by ten layers of silicon strips. 
This inner tracking system is surrounded by the
electromagnetic crystal and the hadronic brass sampling calorimeter.
A superconducting solenoid with a magnetic field of 3.8~T is surrounding the hadronic 
calorimeter with its purpose to bend charged particle tracks, which in turn allows
the determination of their momenta once the particle type is identified. 
Outside of the magnet are located four layers of gaseous muon chambers, embedded between the
iron return yoke to focus the magnetic flux inside the detector volume.
Similar detector technologies are implemented in the endcaps in form of disks
for hermetic coverage.

The different subsystems of the CMS detector are operational to 98\% in average 
(see Fig.~\ref{fig0:CMSperformance}, left), ranging from
the highest data taking efficiency of nearly 100\%
reached by the hadronic calorimeter, to the lowest data taking efficiency
of 96.3\% of the silicon strips. Altogether there are bout 80 million channels to be read out.
The high efficiency helps to keep the reconstruction and the comparison to simulation simple.

The rate of $pp$ collision produces a data volume corresponding to about 2~TB per second. To reduce this huge amount of data to a manageable size, a Level-1 trigger based on fast electronics implemented in programmable chips is employed. With an acceptance 
rate of up to 100~kHz the data is then send to a computing farm at CMS
with 1600 multicore Linux PC's, where the events are scrutinised with more 
sophisticated algorithms to make a final decision at this 
High Level Trigger~\cite{triggerTwiki} (HLT),
if a given event might be of interest or can be discarded. The HLT accept rate
can reach values closely above 1000~Hz. The rate of the most important triggers 
at HLT level agree with simulation within 5\%.
Over 300~TB local disk space permits the 
temporary storage of the data for up to one week. In normal operation the data is send
continuously to the CERN computing centre for storage, reconstruction and distribution to computing centres around the world.
An overview of the CMS infrastructure and data flow is 
depicted in Fig.~\ref{fig0:CMSperformance}, right.

In 2011 up to 130~pb$^{-1}$ of integrated luminosity per day have been 
recorded~\cite{lumiPAS}\cite{lumiTwiki}.
Peak luminosities up to $37 \cdot 10^{32}$~Hz/cm$^2$ have bean reached and 
an integrated luminosity of 5.7~fb$^{-1}$ has been delivered by the LHC. 
The recorded data undergoes a certification process where depending on the
operationality of the various sub-detector systems and the needs for dedicated physics 
analyses two dataset categories are established. 85\% of the data are certified as good
where all sub-detectors have been fully operational. 90\% of the data are certified
as good for muon analyses disregarding calorimeter quality.

The number of parasitic $pp$ collisions (referred to as pileup events) 
accompanying a high energetic scattering of 
interest in a single LHC beam crossing is distributed according to a Poisson statistics 
with an average number of nine to ten events in 2011 and up to 25 events expected in 
2012. In general the distinct vertices can be reconstructed to sort out interesting 
high energetic physics events from additional low energetic scattering background.
The number of pileup events increases with the number of instantaneous luminosity.
A compromise is made between high luminosity and clean events for an optimised 
discovery potential.

At present proton proton collisions take place at the LHC with a centre of mass
energy of $\sqrt{s}=8$~TeV with a proton bunch spacing of 50~ns. 
Data at previous centre of mass energies of $\sqrt{s}=0.9$, 2.76 and
7~TeV with the same bunch spacing have been analysed and are presented here. 
Dedicated run periods with heavy ion collisions
at a centre of mass energy of 2.75~TeV per nucleon do also take place.
One of the primary goals why the LHC has been built is the discovery of new physics,
in particular the electroweak symmetry breaking mechanism, also referred to as 
Brout-Englert-Higgs (BEH) mechanism which has been primarily conceived to endow 
the particles with mass. 
Experimental analysis of this mechanism will allow to probe the mathematical consistency
of the Standard Model (SM) of particle physics beyond the TeV energy scale.
The exploration of new energy frontiers at collider based experiments opens new possibilities
on the way toward a unified theory. Potential discoveries could reveal the existence of
supersymmetry, extra dimensions or dark matter at accessible energy scales.
Parallel to the searchs for new particles and phenomena a rich programme of SM physics
in various domains, like multijet production, electroweak vector boson production,
heavy flavour physics and top quark physics are conducted. 


Exclusively results based on data taken in $pp$ collisions until end of 2011 are 
shown in the following.

\begin{figure}[b]
\vspace*{-70ex}
\resizebox{1.38\columnwidth}{!}{%
 \hspace*{-4ex} \includegraphics{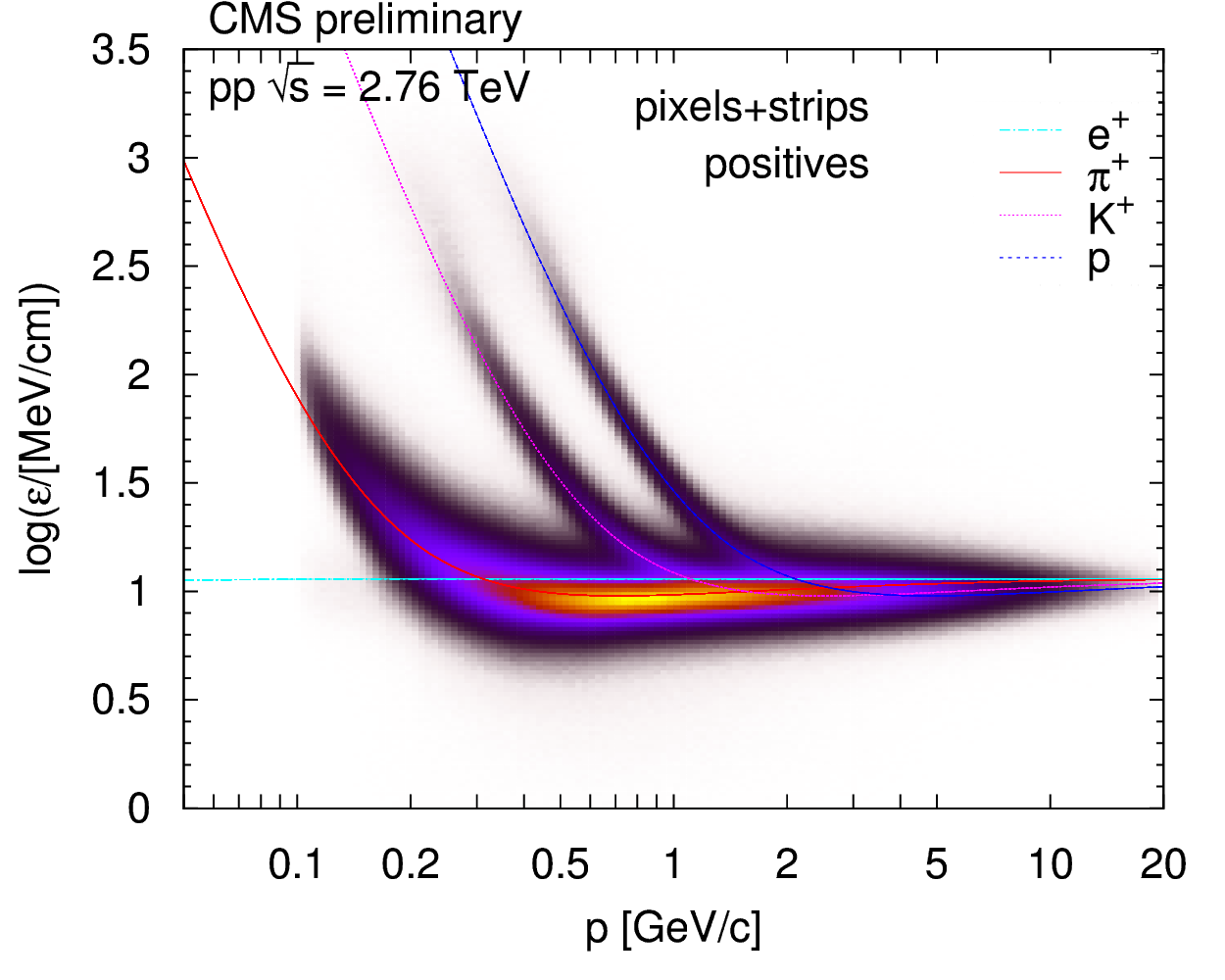} 
 \hspace*{-70ex} \includegraphics{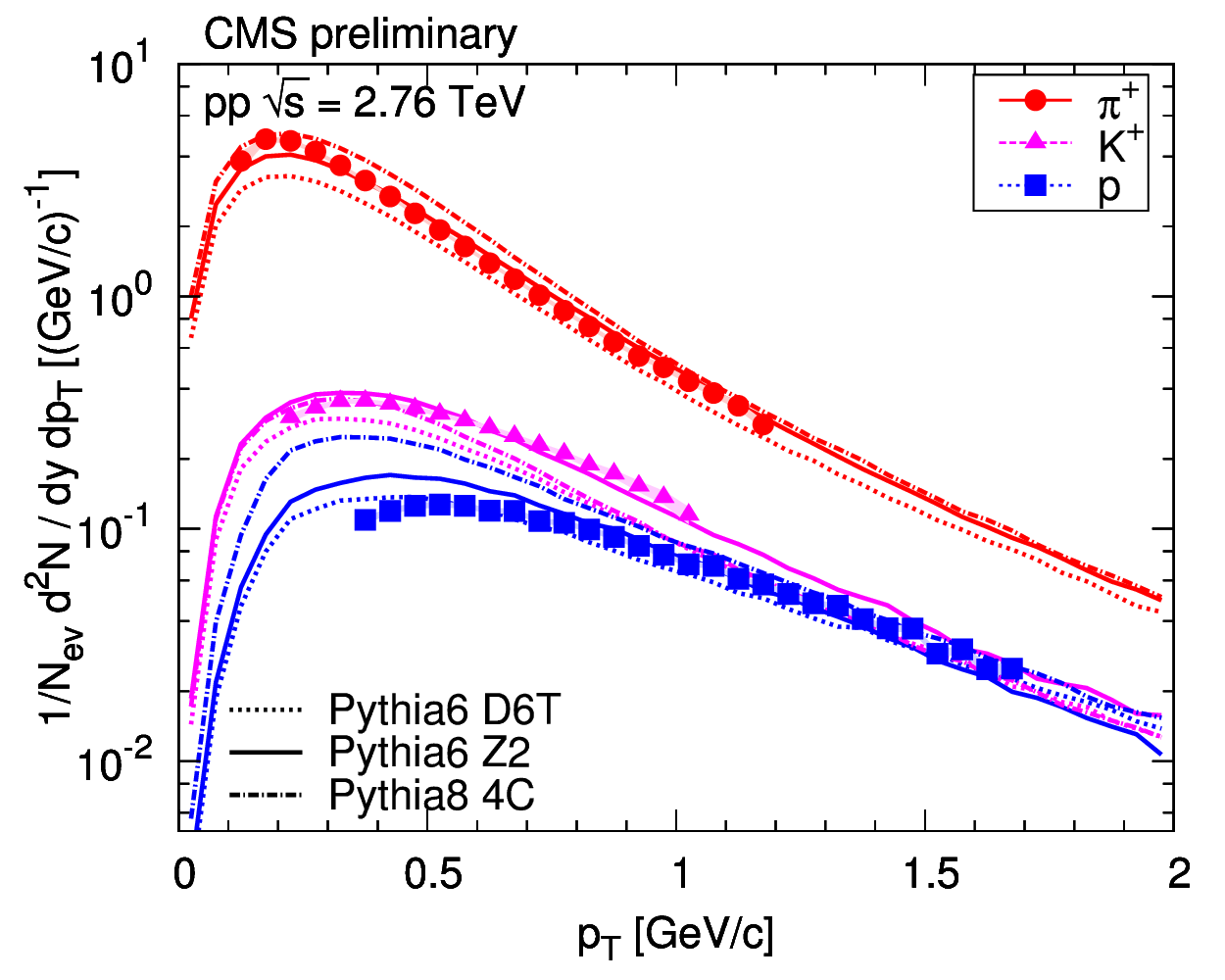} 
 \hspace*{-4ex}
}
\caption{ \label{fig1:dEdx_vs_p}
Logarithmic energy loss in the silicon tracking system as a function of particle momentum (left) and transverse momentum distribution of the different particles (right) in $pp$ collisions at a centre of mass energy of $\sqrt{s}=2.76$~TeV.}
\end{figure}

\section{Particle identification}
Particles are reconstructed and identified using a particle flow algorithm which 
combines the information of the different subdetector systems to distinguish
charged and neutral hadrons (most prominently pions, kaons, protons and neutrons), 
photons, electrons, muons. The reconstructed and identified particles are then used
to construct jets with an anti-$k_T$ algorithm~\cite{antikT}\cite{fastJET} 
(with a distance measure of $R=0.5$ in general), missing transverse energy 
$\not\!\!E_T$ and hadronically decaying (one- and three-prong) taus.

The momentum, transverse momentum and particle multiplicity spectra of identified 
charged hadrons are measured in $pp$ collisions at centre of mass energies of 
$\sqrt{s}=0.9, \; 2.76$ and 7~TeV~\cite{ParticleIDpas}. Charged pions, kaons and 
protons are distinguished by means of energy loss in silicon tracks and the 
consistency of track fits. 
The corrected spectra are compared to various underlying 
event tunes and event generators. 
In Fig.~\ref{fig1:dEdx_vs_p} (left) the logarithmic 
energy loss in the silicon tracking system as a function of the particle momenta
is shown for different charged particles. The right plot shows the particle
multiplicity spectrum as a function of the particle transverse momentum.
The average $p_T$ of pions, kaons and protons 
increases rapidly with the particle mass and the event charged particle multiplicity.
No dependence of the hadronic centre of mass energy can be observed.
The average $p_T$ of protons can not be reproduced by any model.

\begin{figure}[b]
\vspace*{-5ex}
\resizebox{0.85\columnwidth}{!}{%
 \hspace*{-19ex} \includegraphics{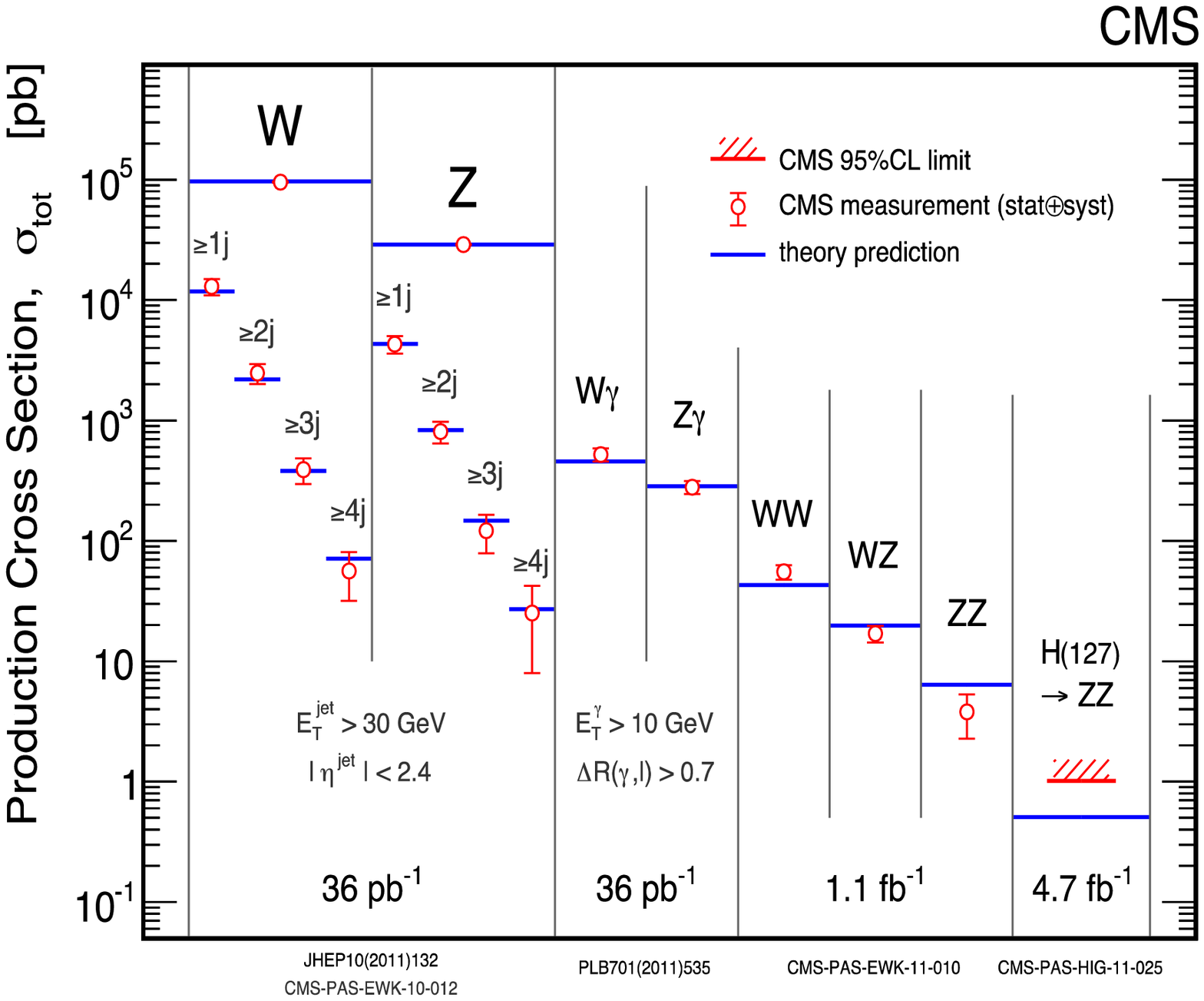} 
 \hspace*{95ex}
}
\vspace*{-46ex}
\unitlength 1cm
\begin{picture}(10., 3.)
 \put(5.92,6.32){ \includegraphics[width=7.4cm]{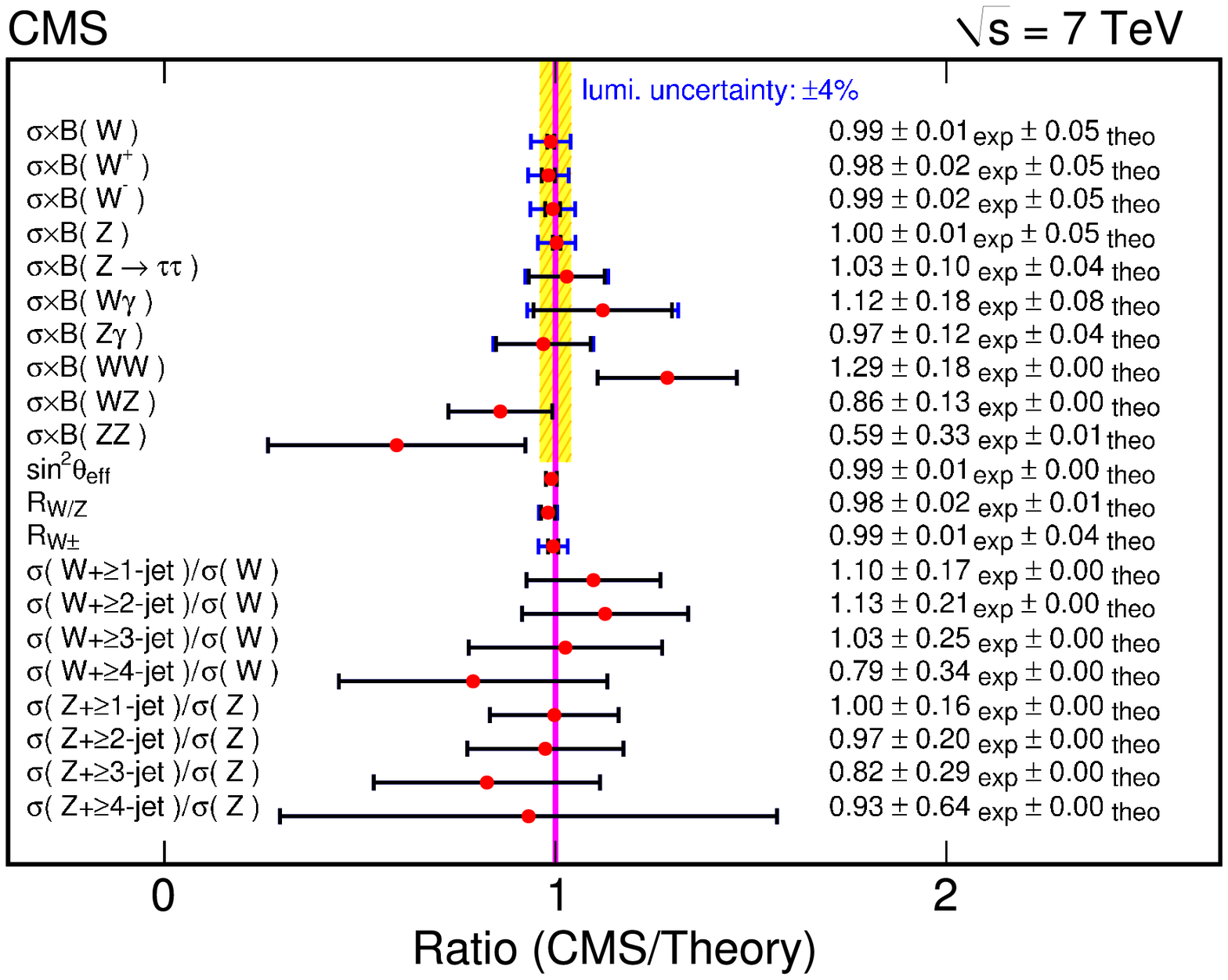} }
\end{picture}

\caption{ \label{fig2:vecBosXsecs}
Various vector boson cross section measurements (left) in comparison to theory 
prediction and the ratio of electroweak observables (right) with respect to the 
theory prediction. 
}
\end{figure}

\section{Standard Model physics} \label{sec:3}
The well known SM particles which have been originally discovered over the
last century, starting with the discovery of the muon in 1933, followed by the pion,
kaon, hyperon, $J/\psi$, $\Upsilon$, the $W$ and $Z$ vector bosons and the
top quark in 1995, have all been rediscovered at CMS between 2006 and 2011.
This demonstrates the well understood performance of the CMS detector, trigger
and software.

Inclusive differential jet cross sections have been measured and Parton Density 
Function (PDF) constraints set in inclusive jet production~\cite{CMS-PAS-QCD-11-004}. 
Transverse jet momenta extend up to 2~TeV and invariant dijet masses up to 5~TeV.
The data are corrected for detector effects to obtain the hadronic final state
and they are compared to perturbative QCD at Next-to-Leading-Order (NLO) augmented by 
non-perturbative corrections to avoid phenomenological model dependence of the 
corrected data. Good agreement is observed between data and the prediction from 
theory.

Another important aspect of SM physics can be tested in the sector of
electroweak physics by means of vector boson production.
The recent measurement of the Drell-Yan differential and double differential cross 
section~\cite{CMS-PAS-EWK-11-007} in the electron and muon channels as a function 
of the invariant dilepton mass
and the rapidity $|y|$ of the dilepton system demonstrates good agreement with 
respect to NNLO~\cite{fewz} predictions and proton PDF's.

The first observation of $Z\rightarrow 4\ell$ production has been 
made~\cite{CMS-PAS-SMP-12-009}. It constitutes an important background process
for $H\rightarrow ZZ\rightarrow 4\ell$ production. The second lepton pair can be 
produced via a virtual photon from radiation in the initial or final state.
The branching ratio has been measured to be 
$BR(Z\rightarrow 4\ell)=[4.4^{+1.0}_{-0.8}(\mbox{stat})+0.2(\mbox{syst})]\times 10^{-6}$ 
in good agreement with the SM prediction of $4.45\times 10^{-6}$.

\begin{figure}[b]
\vspace*{-45ex}
\resizebox{1.0\columnwidth}{!}{%
 \hspace*{-8ex} \includegraphics{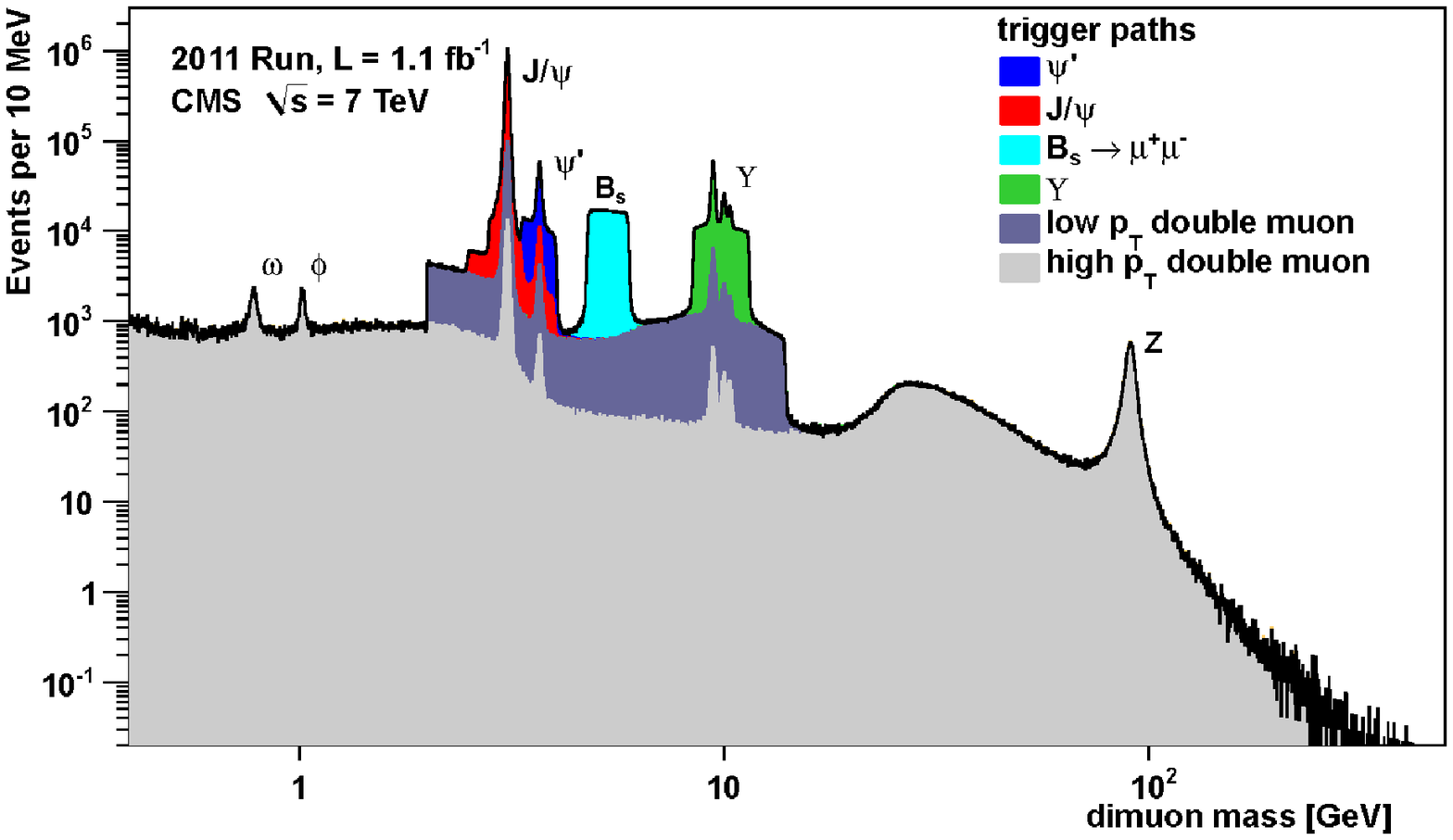} 
 \hspace*{90ex}
}
\vspace*{-42ex}
 \unitlength 1cm
 \begin{picture}(10.,6.)
 \put(7.1,10.7){ \includegraphics[width=6.2cm,angle=270]{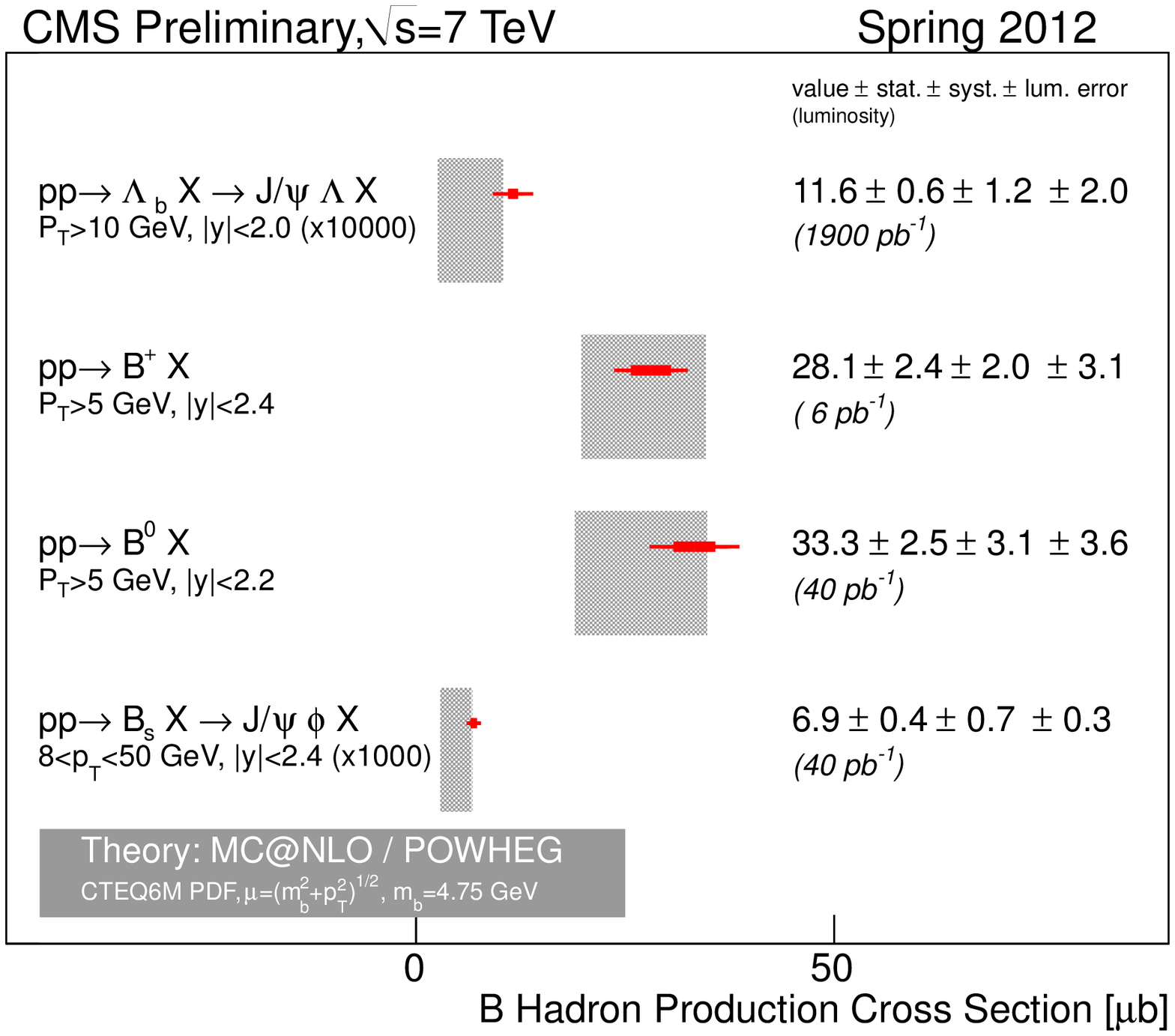} }
 \end{picture}
\caption{ \label{fig3:BtrigAndXsecs}
Dimuon invariant mass distribution (left) collected with various dimuon triggers. 
making use of various low and high $p_T$ thresholds.
On the right shown is a summary of CMS $B$ hadron cross section 
measurements in comparison to theory prediction at NLO accuracy.
}
\end{figure}

The Drell-Yan forward backward asymmetry $A_{FB}$ has been 
measured~\cite{CMS-PAS-EWK-11-004} in different $|y|$ bins as a function of the 
invariant dilepton mass. $e^+e^-$ and $\mu^+\mu^-$ channels have been combined.
It is the first time that this observable has been measured in the $\mu^+\mu^-$
channel at a hadron collider. The measured distributions are unfolded to the 
Born level and in good agreement with the SM prediction.

Various vector boson cross section measurements~\cite{CMS-PAS-EWK-10-012}\cite{PLB701_2011_535}\cite{CMS-PAS-EWK-11-010}
as well as exclusion limits for the production of a scalar boson\cite{CMS-PAS-HIG-11-025} have been established as depicted in Fig.~\ref{fig2:vecBosXsecs} (left plot)
encompassing vector boson plus inclusive jet production, vector boson plus photon
production, diboson production and search for $H\rightarrow ZZ$. The right plot
shows the CMS data over theory ratio of many electroweak observables indicating agreement
between measurement and SM theory predictions.

\section{Heavy flavour physics}
A typical signature of $B$ hadron decays is the production of muons.
In particular muon pair production is of interest with respect to
neutral resonance decays like $J/\psi$, $B_{s(d)}^0$, $\Upsilon$, etc.
Dedicated dimuon triggers are implemented to improve the statistics 
and thus the significance of relevant analysis channels.
The invariant dimuon mass distribution is shown in Fig.~\ref{fig3:BtrigAndXsecs}, left.
The data has been collected with various dimuon triggers 
during the first 1.1~fb$^{-1}$ of data taking in 2011. 
The coloured areas correspond to dimuon triggers with low transverse momentum ($p_T$)
thresholds collected in narrow mass windows. The light gray continuous distribution 
represents events collected with a high $p_T$ threshold dimuon trigger. 
The dark gray band indicates a "quarkonium" dimuon trigger employed during the 
first 220~pb$^{-1}$ of 2011 data.

A new baryon resonance $\Xi_b^{*0}$ containing a $b$ quark flavour has been 
observed~\cite{CMS-PAS-BPH-12-001} via its strong decay into $\Xi_b^{\pm}\pi^{\mp}$.
The known $\Xi_b^{\pm}$ baryon is reconstructed via the decay chain
$\Xi_b^\pm\rightarrow \, J/\psi\Xi^\pm$ with $J/\psi\rightarrow \mu^+\mu^-$ and
$\Xi^\pm\rightarrow \Lambda^0\pi^\pm$ whereas the $\Lambda^0$ in turn is reconstructed
via its decay into $p\pi^-$ (and optionally charge conjugated 
$\bar{\Lambda}^0\rightarrow\bar{p}\pi^+$).
A peak is observed in the distribution of the difference between the mass of the
$\Xi_b^\pm\pi^\mp$ system and the scalar sum of the masses of the
$\Xi_b^\pm$ and the $\pi^\mp$, with a statistical significance of 6.9 standard 
deviations. The mass difference of the peak corresponds to
$14.84 \pm 0.74(\mbox{stat}) \pm 0.28(\mbox{syst})$~MeV. 
Most likely the new state corresponds to the $J^P = 3/2^+$ companion of the 
$\Xi_b^\pm$.

A summary of $B$ hadron production cross section measurements performed with the
 CMS experiment in $pp$ collisions at a centre of mass energy of $\sqrt{s}=7$~TeV
is given in Fig.~\ref{fig3:BtrigAndXsecs}, right plot. The inner error bars of the 
data points correspond to the statistical uncertainty, while the outer (thinner) 
error bars correspond to the quadratic sum of statistical and systematic 
uncertainties. The outermost brackets indicate to the total error, 
including a luminosity uncertainty added in quadrature. 
The measurements are  in good agreement with theory predictions at NLO,
obtained by means of MC@NLO~\cite{mc@nlo1}\cite{mc@nlo2}. 

A search for the rare decays $B^0_{s(d)}\rightarrow\mu^+\mu^-$ is 
performed~\cite{CMS-PAS-BPH-11-020} making use of an integrated luminosity 
of 5~fb$^{-1}$.
The decays into a charge conjugated muon pair are highly suppressed in the SM:
$BR(B_s^0\rightarrow \mu^+\mu^-)=[3.2\pm0.2]\cdot 10^{-9}$ and
$BR(B_d^0\rightarrow \mu^+\mu^-)=[1.0\pm0.1]\cdot 10^{-10}$.
This decay mode is indirectly sensitive to new physics. For example in the
MSSM the branching ratio is proportional to the sixth power of the
vacuum expectation ratio $\tan\beta$.
The analysis exploits a low $p_T$ dimuon trigger and is done blind, making use
of the decay $B^+\rightarrow J/\psi K^+$ for normalisation.
Efficiency is determined in control regions by means of the decay mode
$B^0\rightarrow J/\psi\phi$.
In both decays, $B_s^0$ and $B_d^0$, the number of events observed after all 
selection requirements is consistent with the expectation from background 
plus standard model signal prediction. The resulting upper limits on the branching 
fractions at 95\% Confidence Level (C.L.) are 
$BR(B_s^0 \rightarrow \mu^+ \mu^-) < 7.7\cdot 10^{-9}$ and 
$BR(B_d^0 \rightarrow \mu^+ \mu^-) < 1.8\cdot 10^{-9}$.
A combination of the CMS results with those of the ATLAS and LHCb experiments
improves the upper limits on the branching fractions to 
$BR(B^0_s \rightarrow \mu^+ \mu^-) < 4.2 \cdot 10^{-9}$ and 
$BR(B^0 \rightarrow \mu^+ \mu^-) < 8.1 \cdot 10^{-10}$, both at 95\% C.L. 

In~\cite{Straub2012} CMS and LHCb limits are exploited to put severe constraints
on allowed phase space of supersymmetry and other theories. Most Beyond the SM
(BSM) extensions are able to cope with SM $BR(B^0_{s(d)}\rightarrow\mu^+\mu^-)$ but
new physics could have well shown up already.

\begin{figure}[t]
\vspace*{-4ex}
\resizebox{0.38\columnwidth}{!}{%
 \hspace*{-12.5ex} \includegraphics{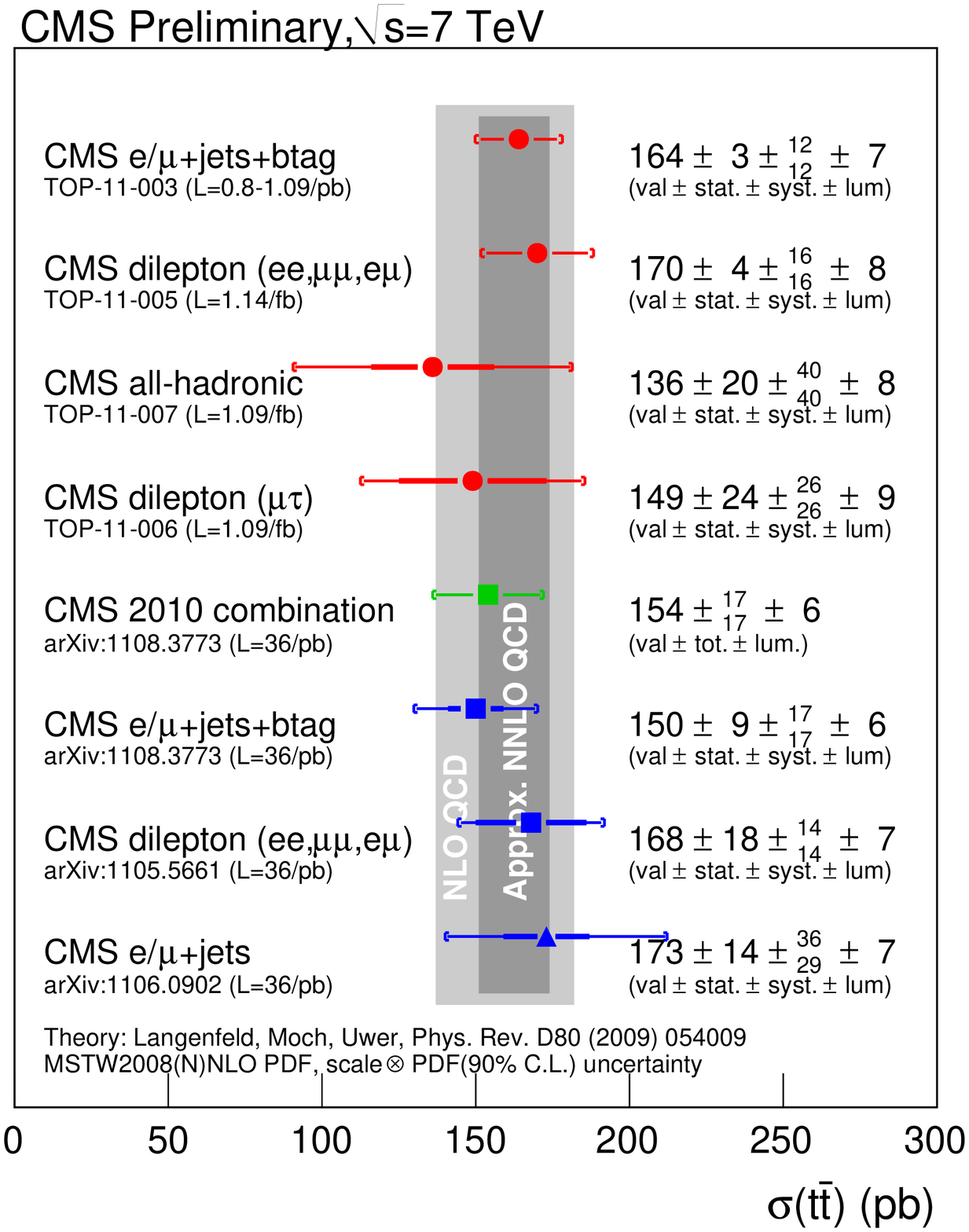} 
 }
 \hspace*{-40ex} 
 \unitlength 1cm
 \begin{picture}(10., 3.)
 \put(0.12,-17.06){ \includegraphics[width=21.2cm]{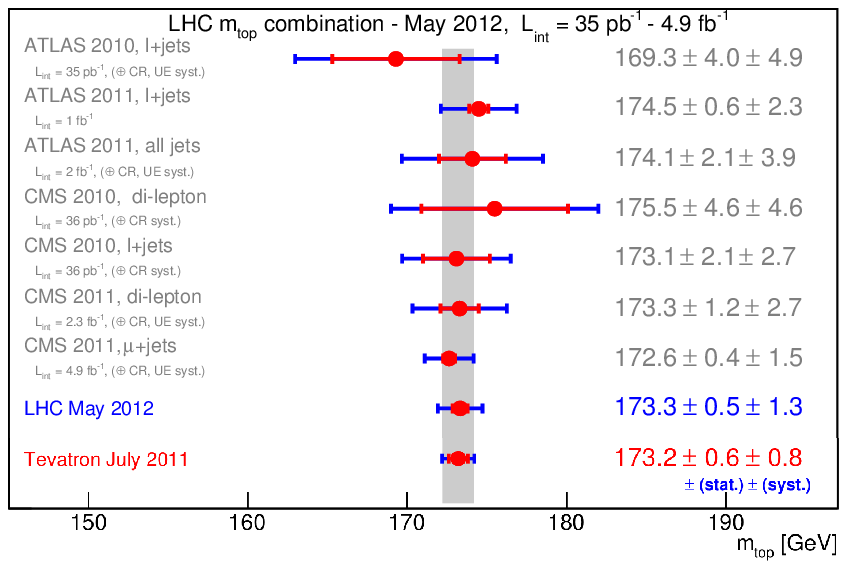} }
 \end{picture}

\caption{ \label{fig4:top_summary}
For $pp$ collisions at $\sqrt{s}=7$~TeV: $t\bar{t}$ cross sections (left) of various decay channels and the combination based on 2010 data and 
LHC top mass determinations and combination (right). 
}
\end{figure}

\section{Top quark physics}
The top quark is the heaviest known particle to-date and it is produced 
pair-wise or singly with associated particles.
The $t\bar{t}$ production at the LHC in $pp$ collisions is dominated by gluon 
gluon fusion (~90\%), the remainder being due to quark antiquark annihilation 
(~10\%). At a centre of mass energy of $\sqrt{s}=7$~TeV the NLO cross 
section~\cite{mc@nlo1} yields 158~pb and the approximated NNLO cross 
section~\cite{ttbarNNLO}
163~pb. In the SM the top quark decays almost exclusively into a $W$ boson and a $b$
quark. The $W$ boson in turn decays in 2/3 of cases into a $q\bar{q}$ pair
and in 1/3 of cases leptonically. Single top quark production is dominated by
the $t$-channel with a cross section of 64~pb at $\sqrt{s}=7$~TeV. Followed by the
$tW$-channel with 15.6~pb and the $s$-channel with 4.6~pb. 

The $t\bar{t}$ cross section measurements~\cite{CMS-PAS-TOP-11-024} are done in all 
possible decay channels by means of a binned likelihood fit.
The left plot of Fig.~\ref{fig4:top_summary} shows the
different decay channel measurements in comparison to theory. 
The dilepton channel is treated as a counting experiment. The hadronic channel 
making use of an unbinned likelihood is taken as one bin.
The Best Linear Unbiased Estimate (BLUE) method~\cite{blue} is used as a cross 
check for the combination of the different channels taking into account 
correlations between different contributions to the measurements.
The combined cross section yields 
$\sigma(t\bar{t})=165.8\pm 2.2(\mbox{stat})\pm10.6(\mbox{syst})\pm 7.8(\mbox{lumi})$~pb
in good agreement with theory at NLO and approximative NNLO order. The relative
error of the cross section $\delta\sigma/\sigma$ amounts to 8\%.

The $t\bar{t}$ production cross sections in $pp$ and $p\bar{p}$ collisions
are also compared to theory at NLO and approximative NNLO order as a function of 
the hadronic centre of mass energy~\cite{CMS-PAS-TOP-11-001}.
Good agreement between the measurements and the prediction from theory is observed.

Single top quark production has been measured~\cite{CMS-PAS-TOP-11-021}
in the $t$-channel with the $W$ boson decaying leptonically as 
$\sigma=70.2\pm 5.2(\mbox{stat})\pm 10.4(\mbox{syst})\pm 3.4(\mbox{lumi})$~pb.
In the $tW$-channel a signal significance of 2.7 standard deviations has been
determined. The $t$-channel single top production is also being exploited
for a direct measurement of the CKM matrix element 
$|V_{tb}|=\sqrt{\sigma_{t}/\sigma^{\mbox{\scriptsize theo}}_{t}(|V_{tb}\equiv 1|)} = 1.04 \pm 0.09(\mbox{exp})\pm 0.02(\mbox{theo})$.

The top mass combination of CMS 2010 and 2011 results in the dilepton and 
lepton plus jets channels are obtained by means of the BLUE method and yield
$m_{\mbox{\scriptsize top}}=172.6 \pm 0.4(\mbox{stat})\pm 1.2(\mbox{syst})$~GeV.
This result is competitive with Tevatron results. Furthermore a LHC and Tevatron
top mass combination also making use of the BLUE method yields 
$m_{\mbox{\scriptsize top}}=173.3 \pm 0.5(\mbox{stat})\pm 1.3(\mbox{syst})$~GeV~\cite{CMS-PAS-TOP-12-001}. 

\begin{figure}[b]
\vspace*{-19ex}
\resizebox{0.44\columnwidth}{!}{%
 \hspace*{-10.0ex} \includegraphics{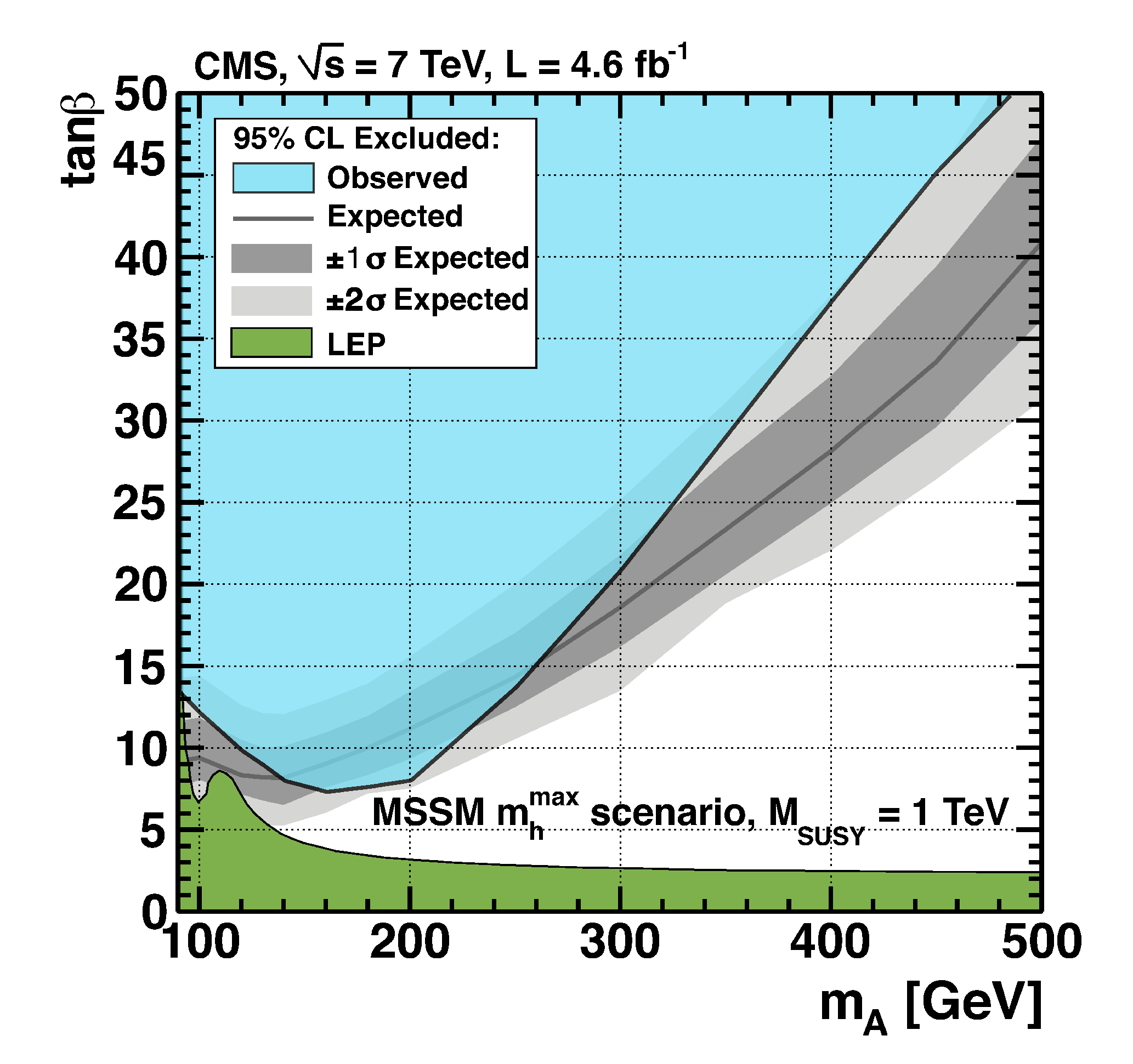} 
 }
 \hspace*{-40ex} 
 \unitlength 1cm
 \begin{picture}(10., 3.)
 \put(4.8,-0.3){ \includegraphics[width=8.3cm]{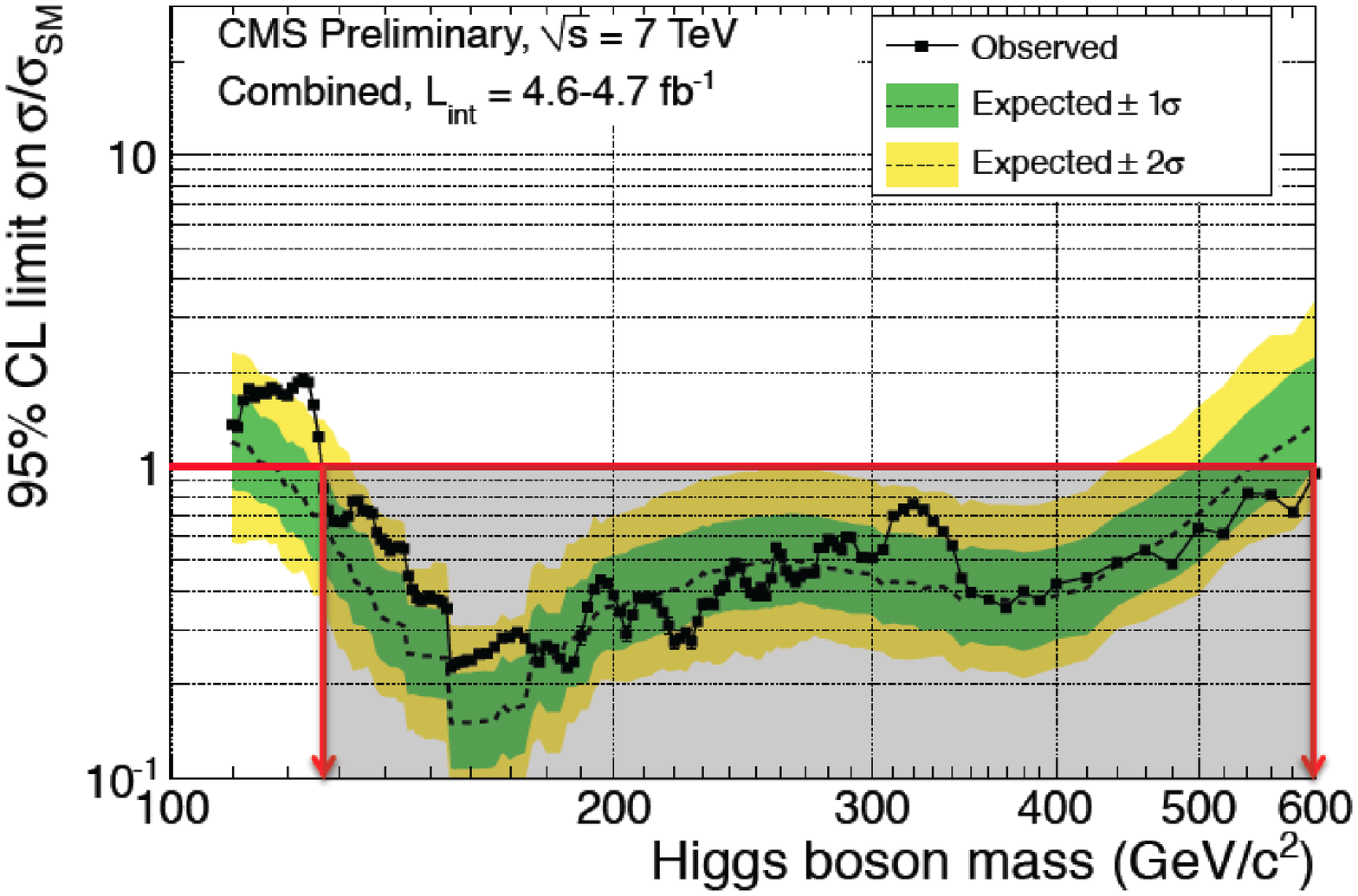} }
 \end{picture}

\caption{ \label{fig5:higgs_limits}
95\% C.L. exclusion limits for a pseudoscalar boson interpreted within the MSSM 
framework in the phase space of $\tan\beta$ as a function of $m_A$ (left) and 
for the scalar SM Brout (alias Higgs) boson as a function of its mass (right). 
}
\end{figure}

\section{Standard Model Brout (alias Higgs) physics}
The electroweak symmetry breaking mechanism, also referred to as 
Brout-Englert-Higgs (BEH) mechanism has been primarily conceived to endow the 
particles with mass. Experimental analysis of this mechanism will allow 
to probe the mathematical consistency of the Standard Model (SM) of particle physics 
beyond the TeV energy scale where divergences of the theoretical description can be 
prevented by means of loop corrections taking contributions and couplings to
a scalar Brout (alias Higgs) boson into account. 
Several production mechanisms contribute at the LHC, namely the dominant gluon fusion
process, the subdominant vector boson fusion and the associated production process.
The associated production via a vector boson suffers a high level of multijet 
background and is less favoured compared to the Tevatron where the reach
is limited to relatively low masses~\cite{Sonnenschein}.
The associated production in $t\bar{t}$ events is too small for consideration
within an integrated luminosity of ${\cal{L}}=5$~fb$^{-1}$.

The Brout boson Yukawa coupling strength is proportional to the mass of the 
particles, favouring the coupling via a top quark loop.
Photons and gluons can only couple indirectly.
The decay modes and branching ratios depend on the Brout mass. Most decay modes are
pursued by CMS.

Electroweak precision measurements by LEP, SLD and Tevatron are sensitive to
the Brout boson mass via radiative corrections.
The $W$ boson (accuracy about 15~MeV) and top quark mass (accuracy about 900~MeV) 
provide constraints as well as electroweak parameter 
fits~\cite{GFitter1}\cite{GFitter2}, indicating a preference for a light mass
of the Brout boson.

The most promising channel in the mass range of $110 < m_H < 150$~GeV
is $H\rightarrow\gamma\gamma$. This channel provides a clean two photon
final state topology. Electromagnetic energy scale and resolution are determined
from $Z\rightarrow ee$, $W\rightarrow e\nu$, $\pi^0$ channels, $E/p$ and
laser calibration for transparency corrections of the crystals of the 
electromagnetic calorimeter accounting for transparency loss due to irradiation.
The corrected energy scale has been kept stable to the 0.1\% level in 2011.
The better are the resolutions the less statistics is needed to achieve the same 
significance of signal over background.
The CMS measurement~\cite{CMS-PAS-HIG-12-001} makes use of multivariate analysis
via Boosted Decision Trees (BDT) for reconstruction of the primary vertex and 
identification of photons and the diphoton system.
About 20\% improvement is obtained compared to a cut based analysis.
Five different significance classes are employed and optimised for expected limits.
Based on an integrated luminosity of ${\cal{L}}=4.76$~fb$^{-1}$ various intervals
in the mass range of $110 < m_H < 150$~GeV are excluded.

A search for a neutral (pseudo-)scalar boson decaying into a pair of taus is 
pursued by CMS~\cite{CMS-PAS-HIG-11-029}. The analysis is sensitive to SM and MSSM 
scenarios. A neutral (pseudo-)scalar boson is produced either directly (via a top 
loop) or via Vector Boson Fusion (VBF) with associated jets. In the case of MSSM
a pseudoscalar boson $\phi$ in association with $b$-jets is produced. 
The direct production suffers from a
large Drell-Yan background. Therefore boosted $\tau\tau$ final states are 
considered in three topologies: $\mu+\tau_h$, $e+\tau_h$ and $\mu e$, making 
use of BDT's in three different SM scenario event categories: 
a) at most one jet giving rise to clean direct production, 
b) boosted boson, giving rise to a high energetic recoil jet
and c) VBF giving rise to two forward jets and rapidity gaps.
In the SM scenario sensitivity can be obtained for $110<m_H<145$~GeV
with the best expected sensitivity at $m_H=120$~GeV.
For the MSSM scenario the associated production of $b$-jets is exploited
in making use of tagging.
$\tan\beta$ can be excluded down to 7.1 at a pseudoscalar boson mass of 
$m_A=160$~GeV. The limits at 95\% C.L. in the phase space of $\tan\beta$
as a function of $m_A$ are shown in Fig.~\ref{fig5:higgs_limits}, left plot. 

The combined CMS Brout (alias Higgs) boson exclusion limits at 
95\% C.L.~\cite{CMS-PAS-HIG-11-032}
making use of the eight analysis channels encompassing the five
decay modes $\gamma\gamma$, $b\bar{b}$, $\tau^+\tau^-$, $W^+W^-$ and 
$ZZ$ are shown in Fig.~\ref{fig5:higgs_limits} right plot. Sensitivity is 
expected between 117 and 543~GeV. The observed limits range from 127 to 600~GeV.
The global significance of observing an excess with a local significance $3.1\sigma$
anywhere in the search range of 110 - 145~GeV is estimated to be $2.1\sigma$.

\begin{figure}[b]
\vspace*{-18ex}
\resizebox{0.38\columnwidth}{!}{%
 \hspace*{-10.0ex} \includegraphics{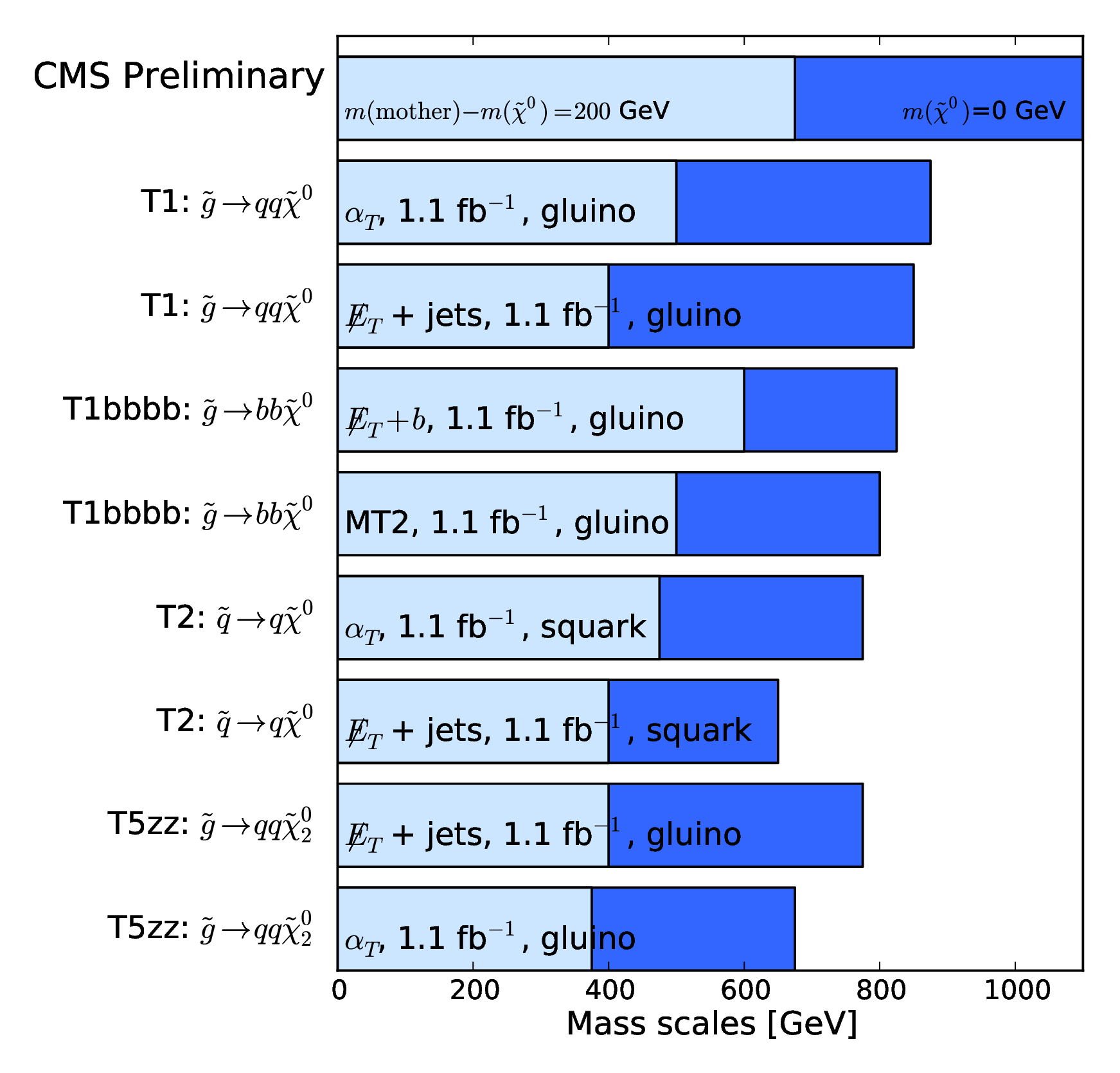} 
 }
 \hspace*{-40ex} 
 \unitlength 1cm
 \begin{picture}(10., 3.)
 \put(4.1,-0.1){ \includegraphics[width=4.0cm]{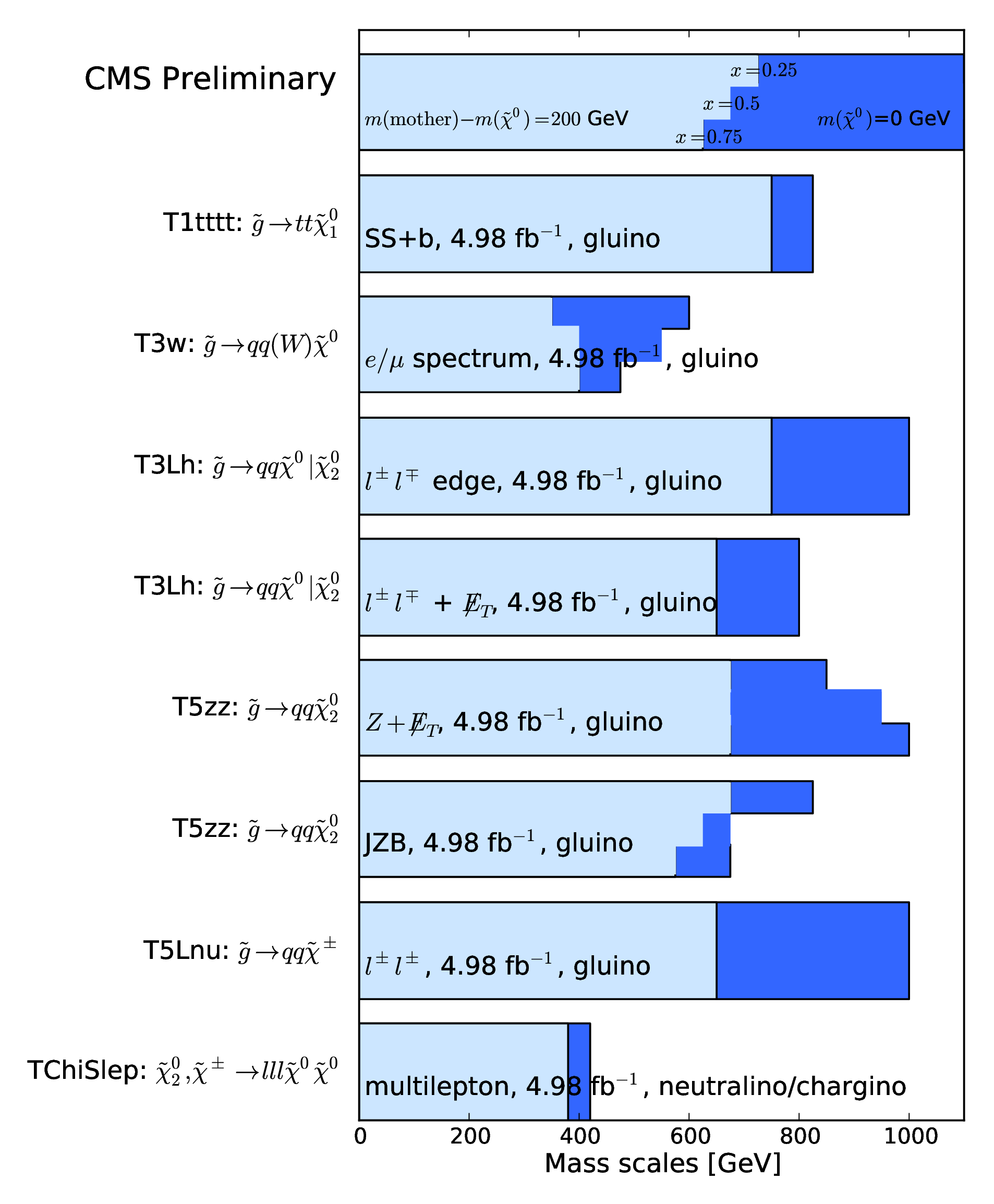} }
 \put(5.2,-1.65){ \includegraphics[width=6.2cm]{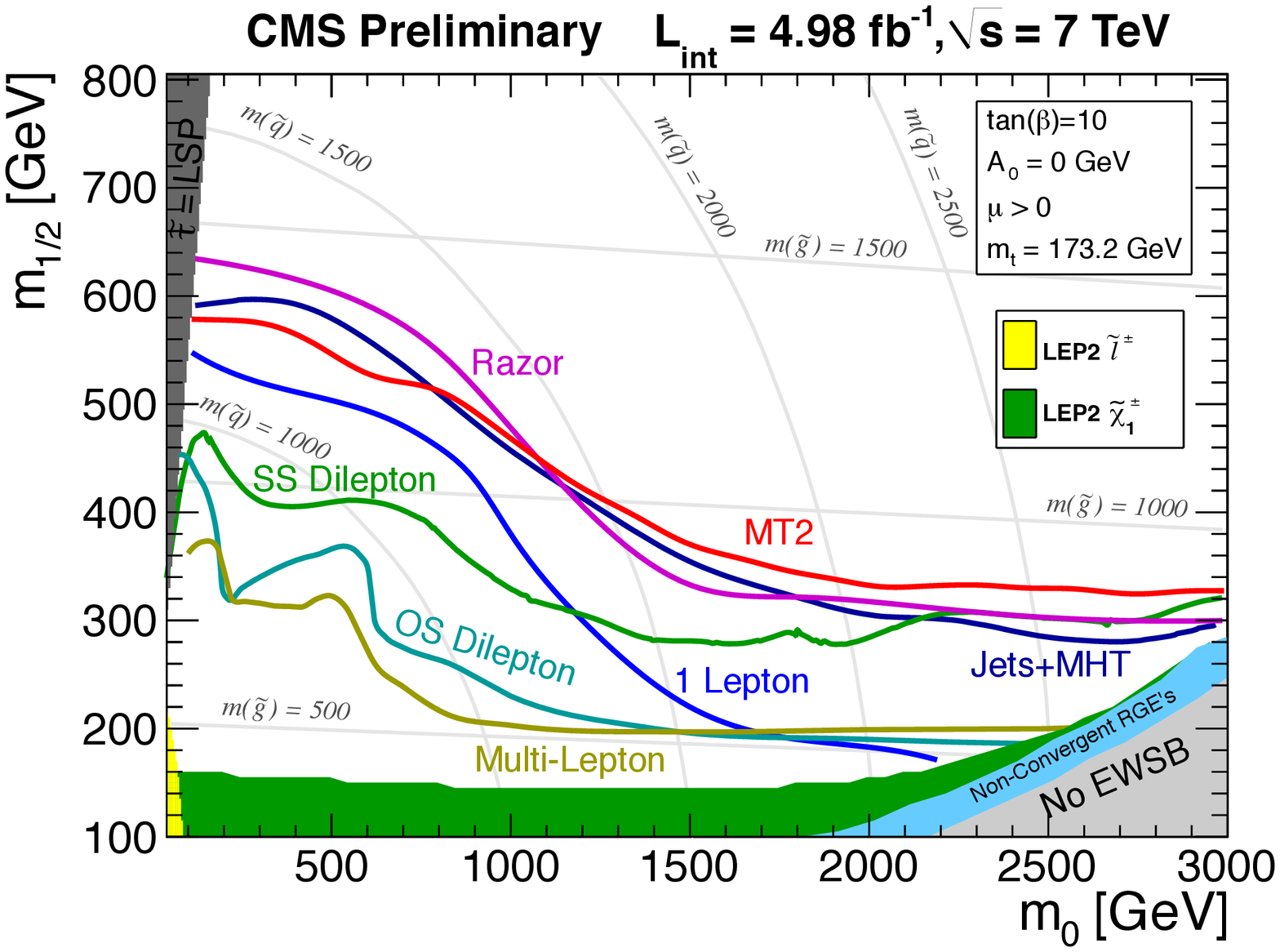} }
 \end{picture}

\caption{ \label{fig6:susy_summary}
Range of excluded mass scale in Simplified Model Spectra (SMS) in hadronic
(left) and leptonic (centre) analysis channels. At the right is shown the
cMSSM exclusion limits summary in the mSUGRA plane of $m_{1/2}$ vs. $m_0$ for 
$\tan\beta=10$, $A_0=0$, $\mu>0$.
}
\end{figure}

\section{Search for supersymmetry}
In the supersymmetric extension of the Standard Model
corresponds to each SM particle a supersymmetric partner particle with identical
gauge quantum numbers but spin differing by 1/2.
The highest expected production cross sections of supersymmetric particles at the 
LHC are given by gluino and squark pair production, followed by the electroweak 
production of charginos and neutralinos.
The decay topology and kinematics is characterised by high transverse momentum jets,
$\not\!\!E_T$ due to the undetected Lightest Supersymmetric Particle (LSP) and
neutralino and chargino decays into SM fermion pairs and the LSP. 

A CMS search for supersymmetry with Razor variables~\cite{CMS-PAS-SUS-12-005} 
exploits an event boost into the Razor frame $\beta_{\mbox{\scriptsize longitudinal}}^{\mbox{\scriptsize Razor}}\equiv (p_z(j_1)+p_z(j_2))/(E(j_1)+E(j_2))$ which corresponds approximately 
to the longitudinal projection of the partonic centre of mass frame.
New physics signal is characterised by a broad peak in the longitudinal invariant
mass $M_R\equiv\sqrt{(E(j_1)+E(j_2))^2 + (p_z(j_1)+p_z(j_2))^2}$ distribution.
The transverse mass $M^R_T\equiv1/2\sqrt{\not\!\!E_T(p_T(j_1)+p_T(j_2))-\not\!\!\vec{E}_T(\vec{p}_T(j_1)+\vec{p}_T(j_2))}$
is introduced to build the dimensionless variable $R\equiv M^R_T/M_R$
which in turn is exploited to discriminate signal from background in a two-dimensional
likelihood fit in the phase space of $R^2$ vs. $M_R$.
Exclusion limits are derived in the cMSSM framework of minimal SUper GRAvity
(mSUGRA) and exposed in the phase space of gaugino mass $m_{1/2}$ as a function of the
scalar mass $m_0$ assuming $\tan\beta=10$, triliniar coupling $A_0=0$~GeV and sign 
$\mu>0$.

Another search for new physics~\cite{CMS-PAS-SUS-11-019} makes use of the 
jet-$Z$-balance $JZB\equiv|-\sum_{\mbox{\scriptsize jets}}\vec{p}_T|-|\vec{p}(Z)|$
in the event to discriminate between background (predominantly $Z+\mbox{jets}$ and 
$t\bar{t}$) and signal which populates larger values of $JZB$.
Three signal search regions: $JZB> 50, 100, 150$~GeV are established to set limits
on the cross section and on neutralino LSP scenarios.

A search for new physics in same-sign dilepton events~\cite{CMS-PAS-SUS-11-010} is 
looking for two like-sign leptons, jets and $\not\!\!E_T$ in various search regions
of $\not\!\!E_T$ and the scalar sum of hadronic transverse energy $H_T$.
Exclusion limits are derived in mSUGRA phase space and for simplified models
with $pp\rightarrow \tilde{g}\tilde{g}$ production and 
$\tilde{g}\rightarrow q\tau\tilde{\chi}^0_1$ decay described by an effective 
Lagrangian.

A range of excluded mass scales in Simplified Model Spectra (SMS) is 
obtained~\cite{CMS-PAS-SUS-11-016} and shown in Fig.~\ref{fig6:susy_summary} 
left chart for models in hadronic decay channels and centre chart for models 
in leptonic decay channnels. 
Constrained MSSM exclusion limits in the $m_{1/2}$ vs. $m_0$ phase space of mSUGRA
for $\tan\beta=10$, $A_0=0$~GeV and $\mu>0$ are shown in Fig.~\ref{fig6:susy_summary},
right plot for various supersymmetric searches. Squark and gluino masses are excluded
up to the multi-TeV-range depending on the considered phase space point.

\begin{figure}[b]
\vspace*{-22ex}
\resizebox{0.345\columnwidth}{!}{%
 \hspace*{-70.0ex} \includegraphics{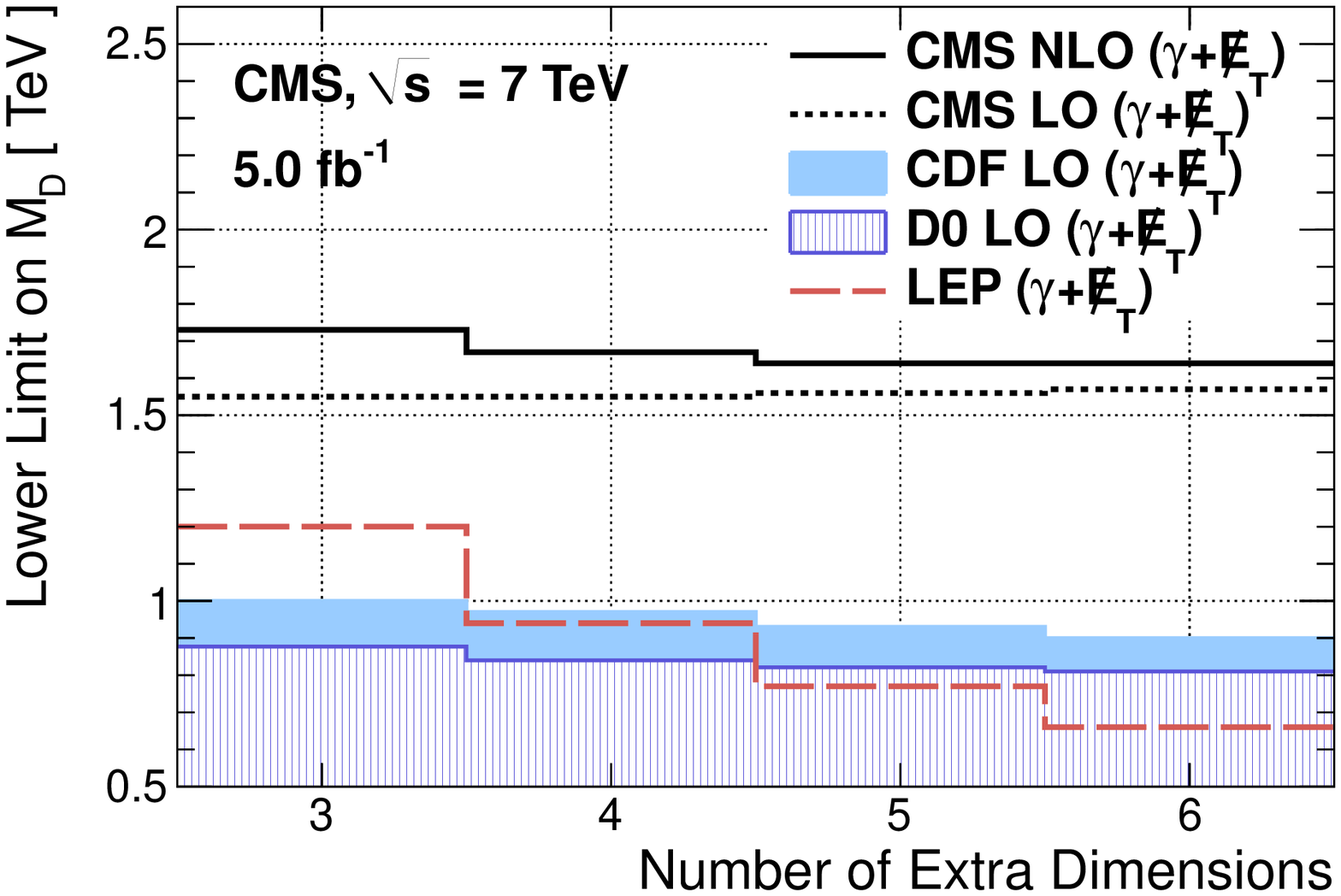} 
 }
 \hspace*{-40ex} 
 \unitlength 1cm
 \begin{picture}(10., 3.)
 \put(7.55, 2.42){ \includegraphics[width=6.8cm]{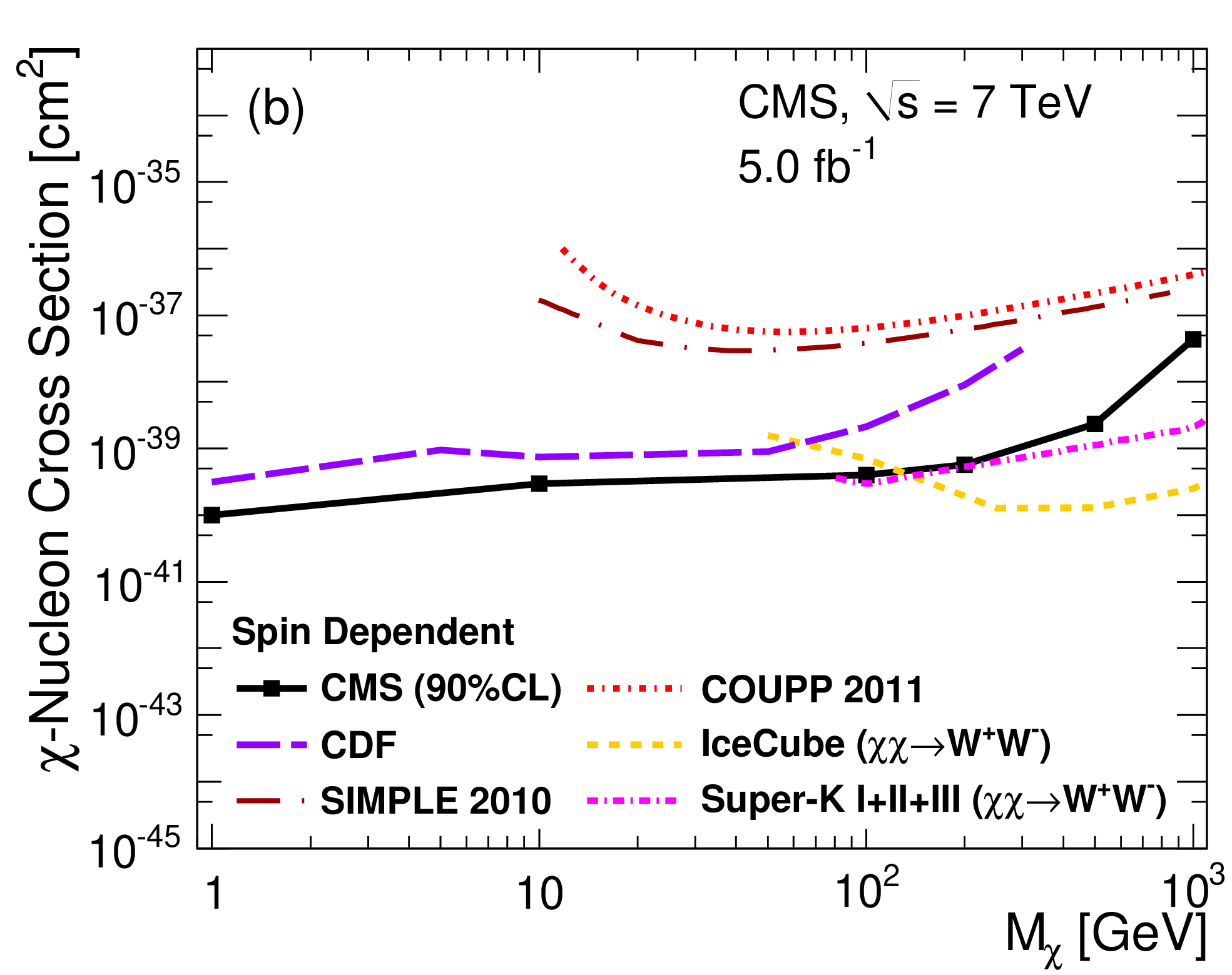} }
 \end{picture}

\vspace*{-18ex}

\caption{ \label{fig7:add_dm}
Limits on the extra-dimensional Planck scale $M_D$ as a function of the number of
Large Extra Dimensions at 95\% C.L. (left) and limits on the Dark Matter
$\chi$-nucleon cross section at 90\% C. L. for comparability with astrophysics 
experiments (right).}
\end{figure}

\section{Search for exotic signatures}
The searches for exotic signatures cover a wide range of models and 
phenomenologies starting with TeV scale gravity. 
The remainder of discussed searches are of too varying 
types and are therefore categorised according to their final state into
searches in lepton, lepton plus jets and jet production. 

A search for Dark Matter (DM) and Large Extra Dimensions (LED) in the
photon plus $\not\!\!E_T$ final state is conducted~\cite{CMS-PAPER-EXO-11-096}.
The ADD~\cite{add} LED model assumes the production process 
$q\bar{q}\rightarrow\gamma G$ where the
graviton $G$ leaves the detector without interaction. 
For models with $3-6$ LED's, extra-dimensional Planck scales between 1.65 and 1.71~TeV
are excluded at 95\% C.L as depicted in Fig.~\ref{fig7:add_dm} (left).
The dark matter model~\cite{DarkMatter} assumes
dark matter pair production $q\bar{b}\rightarrow\gamma D\bar{D}$ where again only 
the photon is expected to be detected. A DM contact interaction scale
$\Lambda(m_{\mbox{\scriptsize mediator}})$ is introduced where spin (in-)dependent
interactions through (vector-) axial-vector couplings are described.
Limits at 90\% C.L. for comparibility with astrophysics experiments are derived.
Cross sections for the production of $\chi$ DM particles in the 
$\gamma$ + $\not\!\!\!E_T$ final state of $13.6 - 15.4$~fb are excluded. 
These are the most sensitive upper limits 
for spin-dependent $\chi$-nucleon scattering for $\chi$ masses $M_{\chi}$ between 1 
and 100~GeV (see Fig.~\ref{fig7:add_dm}). For spin-independent contributions 
present limits are extended to $M_{\chi} < 3.5$~GeV. 

A search for microscopic black holes~\cite{CMS-PAPER-EXO-11-071} is exploiting the
expected energetic multi-particle final state in requiring the scalar event sum 
$S_T$ of transverse momenta of jets, photons, leptons and $\not\!\!E_T$ 
above a certain threshold $S_T^{\min}$ in bins of $N$ final state objects reconstructed
in the event. This discriminating variable is insensitive to Black Hole (BH)
evaporation details. Model independent limits in the multi-TeV range 
are obtained for $3\leq N \leq 8$ by treating the measurement as a counting 
experiment and looking for an excess over SM prediction.
Limits in the multi-TeV range on a Quantum BH mass as function of the 
extra-dimensional Planck scale $M_D$ for $n$ extra-dimensions are obtained, too.
Further limits on a minimum string-ball mass and a semi-classical BH mass
are derived, subject to the constraint that the model approximation breaks down
for $m_{\mbox{\scriptsize BH}}^{\min}\simeq 3-5 M_D$. 

A search for Randall-Sundrum (RS) gravitons in jet plus $\not\!\!E_T$ 
events~\cite{CMS-PAS-EXO-11-061} is looking for the first spin-2 resonance
$R^*$ of Kaluza-Klein modes in the decay channel 
$G^*\rightarrow ZZ \rightarrow q\bar{q}\nu\bar{\nu}$. 
The $Z$ bosons are in general highly boosted. Therefore the event selection
is expecting a collimated $q\bar{q}$ single jet plus $\not\!\!\!E_T$. 
Exclusion limits at 95\% C.L. are set on $\sigma\times BR$ in the range of
0.047 to 0.021~pb for graviton masses of $1000 < M_{G^*} < 1500$~GeV.
Relaxing the perturbative regime of the coupling over Planck scale ratio
from $k/M_{\mbox{\scriptsize Pl}}<0.1$ to $k/M_{\mbox{\scriptsize Pl}}<0.3$
following~\cite{Kelley2011} and translating the cross-section
limits to the parameter space spanned by $M_{G^*}$ and $k/M_{\mbox{\scriptsize Pl}}$, 
limits on the relative coupling parameter $k/M_{\mbox\scriptsize Pl}$ in the range 
of 0.11 to 0.29 are obtained in the considered resonance mass range of 
$1000 < M_{G^*} < 1500$~GeV.
The width of the graviton $G^*$ becomes large for high values of 
$k/M_{\mbox{\scriptsize Pl}}$. A stable signal selection efficiency could be achieved
for $0.05 < k/M_{\mbox{\scriptsize Pl}} < 0.3$.

A search for anomalous production of multilepton events~\cite{CMS-PAPER-EXO-11-045} 
is pursued. Electrons, muons and taus decaying leptonically as well as 
hadronically (1-prong only) are considered in the final state.
$R$-parity is defined as $R_P=(-1)^{3B+L+2S}$ with $B=$ baryon number, $L=$ lepton
number and $S=$ spin. For SM particles holds $R_P=+1$ and for supersymmetric ones
$R_P=-1$. Supersymmetric RPV scenarios
are established by means of an effective Lagrangian
$W_{R\!\!\!\!/\,_P}=\frac{1}{2}\lambda_{ijk}L_iL_j\bar{E}_k + \lambda'_{ijk}L_iQ_j\bar{D}_k + \frac{1}{2}\lambda''_{ijk}\bar{U}_i\bar{D}_j\bar{D}_k$ 
containing $R_P$ violating couplings with generation indices
$i$, $j$, and $k$. $L$ and $Q$ are the lepton and quark $SU(2)_L$ doublet 
superfields and $\bar{E}$, $\bar{D}$, and $\bar{U}$ are the charged lepton, 
down-like quark and up-like quark $SU (2)_L$ singlet superfields, respectively.
The considered scenarios are constrained to short decay lengths of
the $LL\bar{E}$ coupling $L(\lambda_{ijk})\lesssim 100~\mu$m and prompt decays of the
$\bar{U}\bar{D}\bar{D}$ coupling $\lambda_{ijk}''$. 
Only one of the couplings $\lambda_{ijk}$, $\lambda_{ijk}'$ and $\lambda_{ijk}''$
is assumed to be different from zero simultaneously for consistency with the
long proton lifetime.
In particular a leptonic $R_P$ violating scenario (L-RPV) 
with $\lambda_{ijk}\neq 0$, $\lambda_{ijk}' = \lambda_{ijk}=0''$   
and a hadronic  $R_P$ violating scenario (H-RPV) 
with $\lambda_{ijk} = \lambda_{ijk}'=0$ and $\lambda_{ijk}'' \neq 0$
are considered. Exclusion limits at 95\% C.L. on a slepton co-Next-to-Lightest 
Supersymmetric Particle (co-NLSP) scenario with $\lambda_{e\mu\tau}\neq 0$ 
are established in the phase space of $m_{\tilde{g}}$ vs. $m_{\tilde{q}}$, 
exceeding significantly previous searches.
Furthermore limits on H-RPV scenarios with $\lambda_{ijk}''$ are derived.

\begin{figure}[b]
\vspace*{-20ex}
\resizebox{0.32\columnwidth}{!}{%
  \hspace*{-65.0ex} \includegraphics{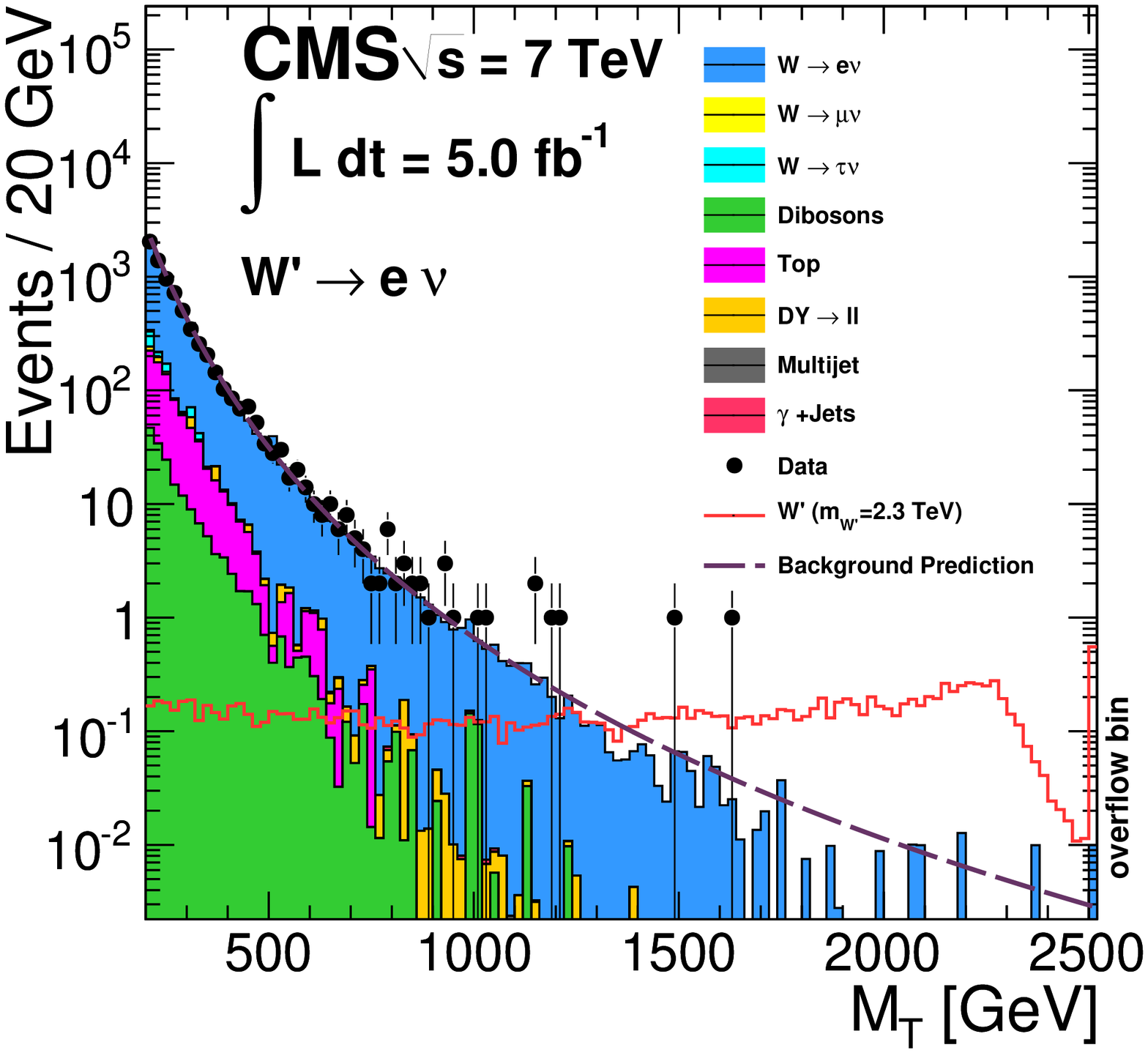} 
}
 \hspace*{-40ex} 
 \unitlength 1cm
 \begin{picture}(10., 3.)
 \put(5.25, 0.3){ \includegraphics[width=6.6cm]{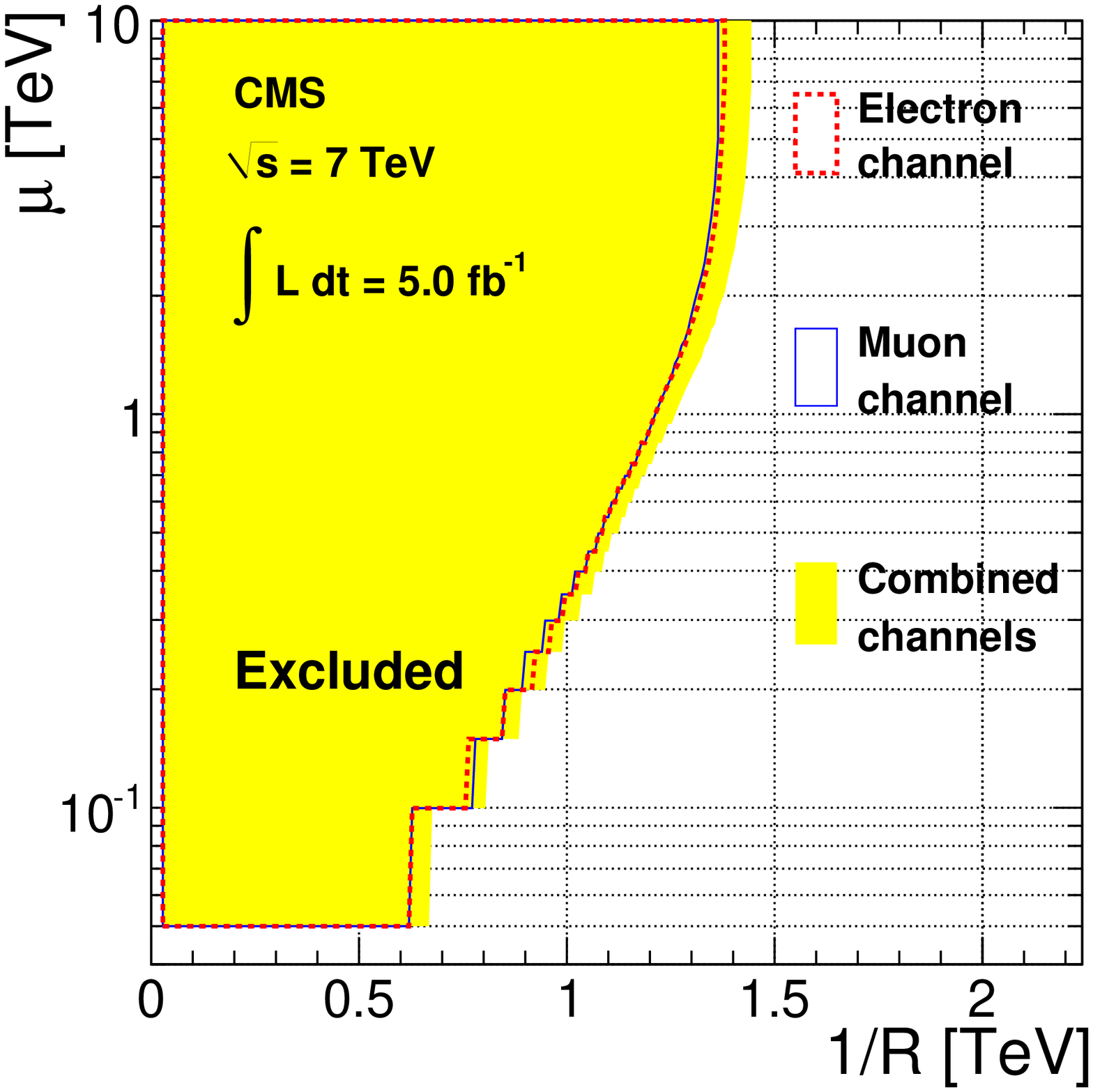} }
 \end{picture}
\vspace*{-3ex}
\caption{ \label{fig8:Wprime}
Transverse mass distribution of the final state electron and $\not\!\!E_T$ 
due to the neutrino, in data compared to SM background (left). 
A potential $W'$ signal with $M_{W'}=3.5$~TeV 
is superposed. To the right exclusion limits of Universal Extra Dimensions 
are derived for the Dirac mass term $\mu$ as a function of the inverse radius 
of the extra dimension.}
\end{figure}

A search for exotic particles decaying to a $WZ$ final state in the letponic decay 
channel~\cite{CMS-PAPER-EXO-11-041} has been accomplished.
Two different model interpretations are investigated.
A sequential SM model with a heavy SM-like charged vector boson
$W'\rightarrow WZ \rightarrow 3\ell + \nu$ and a TechniColor (TC) model
with the technihadrons $\rho_{TC}$ and $\pi_{TC}$ representing bound states of a new
interaction, giving rise to the reaction 
$\rho_{TC}\rightarrow WZ \rightarrow 3\ell + \nu$. 
The $W$ boson can be reconstructed with a two-fold ambiguity in longitudinal 
neutrino momentum. As choice is being taken the smaller $|p_z^{\nu}|$ solution 
which is right in 75\% of cases. A mass dependend cut on the scalar sum
of transverse lepton momenta is applied for improved background reduction.
Exclusion limits at 95\% C.L. are set for a sequential SM $W'$ boson of mass
$M_{W'}=1143$~GeV. Technicolor exclusion limits are set for the techni-rho mass
in the range $167 < M_{\rho_{TC}} < 687$~GeV assuming 
$M_{\pi_{TC}} = 3/4 M_{\rho_{TC}} - 25$~GeV
and in the range $180 < M_{\rho_{TC}} < 938$~GeV assuming $M_{\rho_{TC}} < M_{\pi_{TC}} + M_W$. 

A search for $W'\rightarrow\ell\nu$ with $\ell=e,\mu$ and $\not\!\!E_T$ in the
final state~\cite{CMS-PAPER-EXO-11-024} is interpreted in the sequential SM
and also within the framework of Univeral Extra Dimensions (UED). 
The transverse charged vector boson mass distribution
$M_T=\sqrt{2p_T^{\ell}\cdot \not\!\!E_T  \cdot (1-\cos\Delta\phi_{\ell,\nu})}$
is shown in Fig.~\ref{fig8:Wprime} with good agreement between data and SM
background. Exclusion limits are set at 95\% C.L. for the mass of a SM-like $W'$
boson of 2.5~TeV (right-handed) and 2.63~TeV [2.43~TeV] for constructive 
[destructive] $W - W'$ interference (left-handed). Higher order electroweak corrections
at high masses reduce interference effects. A re-interpretation in a UED model
provides limits in terms of the Extra Dimension (ED) radius $R$ and the Dirac mass 
term $\mu$, whereas no sensitivity to $n\geq 4$ ED modes is obtained yet.

A search for new physics in highly boosted $Z\rightarrow \mu\mu$ 
events~\cite{CMS-PAS-EXO-11-025} is accomplished and interpreted in the context of
an excited quark from gauge and contact interactions in the decay channel 
$q^*\rightarrow qZ$ and $Z\rightarrow \mu\mu$.
Due to the highly boosted $Z$ boson decay the two muons are expected to be collinear.
Therefore the muon isolation check has to take properly 
into account another close-by muon.
Furthermore the invariant dimuon mass has to be consistent with the $Z$ boson mass.
Exploiting the $1/p_T(\mu\mu)$ spectrum a robust Drell-Yan background template 
function can be constructed covering all events up to the highest $p_T(\mu\mu)$.
The relative coupling strengths $f_{SU(2)}=f'_{U(1)}=1$ and $0<f_s^{SU(3)}<1$ are assumed.
The scale where the interaction becomes effective is set to the mass of the excited
quark: $\Lambda = M_{q^*}$. Exclusion limits at 95\% C.L. are obtained on the 
production cross section times branching ratio. 
Under the premisses of $M_{q^*}=\Lambda, f=f'=f_s=1$ mass limits of
$M_{q^*}=1.94$~TeV in gauge production and $M_{q^*}=2.14$~TeV in contact 
interactions are established.
For the contact interaction scenario a large phase space is excluded,
encompassing $f_S=0$ for masses up to $M_{q^*}=2.18$~TeV. This has to be compared with
the H1 limit~\cite{H1_qstar_2009} at $M_{q^*}=252$~GeV. 
 
A search for a heavy neutrino and a right handed $W$ boson is 
pursued~\cite{CMS-PAPER-EXO-11-091}. In a right-left symmetric model
extension of the SM a right handed charged vector boson decays into a muon and a 
heavy neutrino: $W_R\rightarrow\mu+N_{\mu}$ whereas the heavy neutrino in turn
decays into a muon and an off-shell right handed charged vector boson:
$N_{\mu}\rightarrow \mu W^*_R$. The right handed boson decays then finally into a quark
antiquark pair: $W_R^*\rightarrow q\bar{q}$. Therefore the signature of the signal 
consists of two muons and two jets in the final state.
Exclusion limits at 95\% C.L. are set on the production cross section times 
branching ratio for various masses $m_{W_R}$ in the multi-TeV range. Limits are also
obtained in the two dimensional phase space spanned by the heavy neutrino mass 
$M_{N_{\mu}}$ and the charged right handed vector boson mass $M_{W_R}$ assuming
equal couplings in the left and right sector and no alternative decay channels.
Heavy neutrino masses are excluded up to $M_{N_{\mu}}=1.5$~TeV and charged right 
handed vector boson masses up to $M_{W_R}=2.5$~TeV. 

A search for new physics in the dijet angular distribution is 
performed~\cite{CMS-PAPER-EXO-11-017}. The distribution is sensitive to the spin
of the exchanged particle and shows no strong dependence on PDF's. 
At least two jets 
are required to
determine the observable $\chi_{\mbox{\scriptsize dijet}}=\exp(|y_1-y_2|)$ which goes over
in the limit of massless jets to the expression $(1+|\cos\theta^*|)/(1-|\cos\theta^*|)$.
The differential distribution $d\sigma/d\chi_{\mbox{\scriptsize dijet}}$ is flat for 
Rutherford scattering. Quark compositeness reveals through characteristic deviations
from a flat distribution.
The data are corrected for detector effects and compared to NLO theory including
non-perturbative corrections for hadronisation and Multiple Parton Interactions
(MPI). Exclusions limits are set at 95\% C.L. on the production
of different color- and isospin-singlet models with constructive and destructive
interferences between QCD and the contact interactions. Contact interaction scales
$\Lambda$ are excluded up to $14.5$~TeV depending on the model.
The limits have been derived on the reconstructed final state as well as on the
for detector effects corrected hadronic final state. Both methods give consistent
results with respect to each other.

\section*{Conclusions}
The CMS experiment is a multi-purpose detector successfully operated at the LHC
where predominantly $pp$ collisions take place at various centre of mass energies
up to $\sqrt{s}=8$~TeV to-date. 
Results of measurements and searches in $pp$
collisions at $\sqrt{s}=7$~TeV are reported, covering particle identification,
Standard Model physics, heavy flavour physics, top quark physics, Standard Model
Brout (alias Higgs) boson physics, searches for supersymmetry and exotic signatures.  
A vaste phase space has already been probed and still more will be explored.


\begin{thebibliography}{}

\bibitem{cms}
CMS Collaboration, 
JINST {\bf 3} S08004 (2008). 

\bibitem{triggerTwiki}
CMS Collaboration, L1 and HLT approved trigger results (2012), \\
{\tt https://twiki.cern.ch/twiki/bin/viewauth/CMSPublic/L1TriggerDPGResults}.

\bibitem{lumiPAS}
CMS Collaboration,
PAS EWK-10-004 (2010).

\bibitem{lumiTwiki}
CMS Collaboration, Luminosity - public results (2012), \\
{\tt https://twiki.cern.ch/twiki/bin/view/CMSPublic/LumiPublicResults}.

\bibitem{antikT}
M. Cacciari, G. P. Salam, G. Soyez,
JHEP 04 (2008) 063, 
doi:10.1088/1126-6708/2008/04/063.

\bibitem{fastJET}
M. Cacciari, G. Salam, 
Phys. Lett. B 641 (2006) 57,
doi:10.1016/j.physletb.2006.08.037.

\bibitem{ParticleIDpas}
CMS Collaboration,
PAS FSQ-12-014 (2012).


\bibitem{CMS-PAS-QCD-11-004}
CMS Collaboration,
PAS QCD-11-004 (2011).
 
\bibitem{CMS-PAS-EWK-11-007}
CMS Collaboration,
PAS EWK-11-007 (2011).

\bibitem{fewz}
R. Gavin, Y. Li, F. Petriello {\it et al.}, 
Comput.Phys. Commun. 182 (2011) 2388–2403.

\bibitem{CMS-PAS-SMP-12-009}
CMS Collaboration,
PAS SMP-12-009 (2012), submitted to JHEP.

\bibitem{CMS-PAS-EWK-11-004}
CMS Collaboration,
PAS EWK-11-004 (2011), submitted to Phys. Lett. {\bf B}.

\bibitem{CMS-PAS-EWK-10-012}
CMS Collaboration,
PAS EWK-10-012, JHEP10 (2011) 132.

\bibitem{PLB701_2011_535}
CMS Collaboration,
EWK-10-008, arXiv:1105.2758 [hep-ex], Phys. Lett. {\bf B} 701 (2011) 535.  

\bibitem{CMS-PAS-EWK-11-010}
CMS Collaboration,
PAS EWK-11-010 (2011), submitted to JHEP.

\bibitem{CMS-PAS-HIG-11-025}
CMS Collaboration,
PAS HIG-11-025 (2011), CERN-PH-EP-2012-025, submitted to Phys. Rev. Lett.

\bibitem{CMS-PAS-BPH-12-001} 
CMS Collaboration,
arXiv:1204.5955 [hep-ex], Phys. Rev. Lett. 108, 252002 (2012).

\bibitem{mc@nlo1} 
S. Frixione and B.R. Webber, 
JHEP 0206 (2002) 029, arXiv:0204244 [hep-ph].

\bibitem{mc@nlo2}
S. Frixione, P. Nason and B.R. Webber, 
JHEP 0308 (2003) 007, arXiv:0305252 [hep-ph].

\bibitem{CMS-PAS-BPH-11-020} 
CMS Collaboration,
PAS BPH-11-020, CERN-PH-EP-2012-086, arXiv:1203.3976, JHEP 04 (2012) 033.

\bibitem{Straub2012}
D. Straub,
arXiv:1205.6094 [hep-ph] (2012).

\bibitem{ttbarNNLO}
U. Langenfeld, S. Moch, P. Uwer,
Phys. Rev. {\bf D} 80 (2009) 054009.

\bibitem{CMS-PAS-TOP-11-024}
PAS TOP-11-024 (2011).

\bibitem{blue}
L. Lyons, D. Gibaut, P. Clifford, 
Nucl. Instr. and Meth. A, 270 (1988) 110.

\bibitem{CMS-PAS-TOP-11-001}
CMS Collaboration,
PAS TOP-11-001 (2011), arXiv:1108.3773 [hep-ex].

\bibitem{CMS-PAS-TOP-11-021}
CMS Collaboration,
PAS TOP-11-021 (2011).

\bibitem{CMS-PAS-TOP-12-001} 
CMS and ATLAS Collaborations,
PAS TOP-12-001, ATLAS-CONF-2012-095, (2012).

\bibitem{Sonnenschein}
L. Sonnenschein,
Habilitation thesis, \\ 
\scalebox{0.96}{\tt http://www-d0.fnal.gov/results/publications\_talks/thesis/sonnenschein/thesis.pdf}, 
Universit\'e Pierre et Marie Curie - Sorbonne Universit\'es, Paris (2006).

\bibitem{GFitter1} 
H. Fl\"acher, M. Goebel, J. Haller, A. H\"ocker, K. M\"onig, J. Stelzer, 
Eur. Phys. J. C 60, 543 (2009), arXiv:0811.0009 [hep-ph]. 

\bibitem{GFitter2}
M. Baak, M. Goebel, J. Haller, A. H\"ocker, D. Ludwig, K. M\"onig, M. Schott, J. Stelzer, 
submitted to Eur. Phys. J. C, arXiv:1107.0975 [hep-ph].

\bibitem{CMS-PAS-HIG-12-001}
CMS Collaboration,
PAS HIG-12-001 (2012).

\bibitem{CMS-PAS-HIG-11-029}
CMS Collaboration,
PAS HIG-11-029 (2011).

\bibitem{CMS-PAS-HIG-11-032}
CMS Collaboration,
PAS HIG-11-032, arXiv:1202.1488 [hep-ex], Physics Letters B 710 (2012) 26–48.

\bibitem{CMS-PAS-SUS-12-005}
CMS Collaboration,
PAS SUS-12-005 (1012).

\bibitem{CMS-PAS-SUS-11-019}
CMS Collaboration,
PAS SUS-11-019 (2011).

\bibitem{CMS-PAS-SUS-11-010}
CMS Collaboration,
PAS SUS-11-019, arXiv:1205.6615 [hep-ex], subm. to Phys. Rev. Lett. (2012).

\bibitem{CMS-PAS-SUS-11-016}
CMS Collaboration,
PAS SUS-11-016 (2011).

\bibitem{CMS-PAPER-EXO-11-096}
CMS Collaboration,
PAPER EXO-11-096, arXiv:1204.0821 [hep-ex], subm. to Phys. Rev. Lett. (2012).

\bibitem{DarkMatter}
Y. Bai, P. J. Fox and R. Harnik, 
JHEP 12 (2010) 048, arXiv:1005.3797v2.

\bibitem{add}
N. Arkani-Hamed, S. Dimopoulos and G. Dvali, 
Phys. Lett. B 429 (1998) 263, arXiv:hep-ph/9803315.


\bibitem{CMS-PAPER-EXO-11-071}
CMS Collaboration,
PAPER EXO-11-071, arXiv:1202 [hep-ex], JHEP04 (2012) 061.

\bibitem{CMS-PAS-EXO-11-061}
CMS Collaboration,
PAS EXO-11-061 (2011).

\bibitem{Kelley2011}
R. Kelley, L. Randall and B. Shuve, 
JHEP 1102 (2011) 014, arXiv:1011.0728 [hep-ph].

\bibitem{CMS-PAPER-EXO-11-045}
CMS Collaboration,
PAPER EXO-11-045, arXiv:1204.5341 [hep-ex], subm. to JHEP (2012).

\bibitem{CMS-PAPER-EXO-11-041}
CMS Collaboration,
PAPER EXO-11-041, arXiv:1206.0433 [hep-ex], subm. to Phys. Rev. Lett. (2012).

\bibitem{CMS-PAPER-EXO-11-024}
CMS Collaboration,
PAPER EXO-11-024, arXiv:1204.4764 [hep-ex], subm. to JHEP (2012).

\bibitem{CMS-PAS-EXO-11-025}
CMS Collabortaion,
PAS EXO-11-025 (2011).

\bibitem{H1_qstar_2009}
H1 Collaboration, 
Phys. Lett. B678 (2009) 335–343, arXiv:0904.3392 [hep-ex].

\bibitem{CMS-PAPER-EXO-11-091}
CMS Collaboration,
PAPER EXO-11-091, to be subm. to Phys. Rev. Lett. (2012).

\bibitem{CMS-PAPER-EXO-11-017}
CMS Collaboration,
PAPER EXO-11-017, arXiv:1202.5535 [hep-ex], subm. to JHEP (2012).

\bibitem{EXO_mass_limits_summary}
CMS Collaboration, Exotica Public Physics Results (2012), \\
\scalebox{0.81}{\tt https://twiki.cern.ch/twiki/bin/view/CMSPublic/PhysicsResultsEXO\#CMS\_EXO\_Summary\_of\_Mass\_Limits}.



\end{thebibliography}
\end{document}